\documentclass[]{lmcs}
\pdfoutput=1
\usepackage[utf8]{inputenc}

\usepackage{lastpage}
\lmcsdoi{18}{1}{21}
\lmcsheading{}{\pageref{LastPage}}{}{}%
{Nov.~30,~2020}{Jan.~31,~2022}{}

\usepackage{booktabs}

\usepackage{stmaryrd}
\usepackage{mathtools} 
\usepackage{amssymb}
\usepackage{xspace}
\usepackage{tikz}
\usetikzlibrary{backgrounds}
\usetikzlibrary{fit}
\usetikzlibrary{automata}
\usetikzlibrary{calc}
\usetikzlibrary{decorations.pathreplacing,decorations.pathmorphing}
\usepackage[ruled,linesnumbered,vlined]{algorithm2e} 

\SetKwComment{Comment}{$\vartriangleright$  }{} 

\newcommand{\doc}{d}
\newcommand{\alphabet}{\Sigma}

\newcommand{\mspan}[2]{\ensuremath{[#1,#2\rangle}}
\newcommand{\spans}{\mathsf{Spans}}

\newcommand{\vars}{\mathsf{Vars}}
\newcommand{\docs}{\mathsf{Docs}}
\newcommand{\vals}{\dom}
\newcommand{\dom}{\mathbf{D}}

\newcommand{\blank}{\ensuremath{\sqcup}}

\newcommand{\tup}{\mathbf{t}}

\newcommand{\vtup}[1]{{#1}\text{-}\mathsf{Tup}}
\newcommand{\vset}{\text{VSet}}

\newcommand{\df}{\coloneqq}
\newcommand{\join}{\bowtie}

\newcommand{\sel}{\mathbf{P}}

\newcommand{\supp}{\mathsf{supp}}

\newcommand{\rase}{\textsc{RA-Enum}\xspace}

\newcommand{\K}{\srd}
\newcommand{\srd}{\mathbb{K}}
\newcommand{\srplus}{\oplus}
\newcommand{\srtimes}{\otimes}
\newcommand{\srzero}{\overline{0}}
\newcommand{\srone}{\overline{1}}
\newcommand{\boplus}{\bigoplus}
\newcommand{\sr}{(\srd,\srplus,\srtimes,\srzero,\srone)}

\newcommand{\Kp}{\srdp}
\newcommand{\srdp}{\mathbb{K^\prime}}
\newcommand{\srplusp}{\oplus^\prime}
\newcommand{\srtimesp}{\otimes^\prime}
\newcommand{\srzerop}{\overline{0^\prime}}
\newcommand{\sronep}{\overline{1^\prime}}

\newcommand{\srp}{(\srdp,\srplusp,\srtimesp,\srzerop,\sronep)}

\newcommand{\linorder}{\preccurlyeq}

\newcommand{\varop}[1]{\Gamma_{#1}}
\newcommand{\vop}[1]{\mathop{#1{\vdash}}}
\newcommand{\vcl}[1]{\mathbin{{\dashv}#1}}

\newcommand{\weight}[1]{{\mathsf{w}}_{#1}}
\newcommand{\rn}{\rho}
\newcommand{\Rn}[2]{Runs({#1},{#2})}
\newcommand{\e}[1]{\emph{#1}}

\newcommand{\reg}[1]{\ensuremath{\textsf{REG}_{#1}}}
\newcommand{\regk}{\ensuremath{\textsf{REG}_{\K}}}

\usepackage{graphics}
\newcommand{\restrict}{{\ensuremath{\upharpoonright}}}

\newcommand{\repspnrk}[2]{\llbracket{#1}\rrbracket^{#2}}
\newcommand{\repspnrw}[1]{\llbracket{#1}\rrbracket^{\srd}}
\newcommand{\repspnr}[1]{\llbracket{#1}\rrbracket}
\newcommand{\B}{\mathbb{B}}
\newcommand{\true}{\mathsf{true}}
\newcommand{\false}{\mathsf{false}}
\newcommand{\morphtob}{\ensuremath{f}}
\newcommand{\funcfromb}{\ensuremath{g}}

\newcommand{\N}{\mathbb{N}}
\newcommand{\Z}{\mathbb{Z}}
\newcommand{\Q}{\mathbb{Q}}

\newcommand{\np}{\NP}
\newcommand{\NP}{\text{NP}\xspace}
\newcommand{\ptime}{\text{PTIME}\xspace}

\newlength\boxwidth
\newlength\questionwidth
\newcommand{\algproblem}[4]{
  \setlength\boxwidth{\columnwidth}\addtolength\boxwidth{-#1}{
  \setlength\questionwidth{\columnwidth}\addtolength\questionwidth{-#1}\addtolength\questionwidth{-2.5cm}{
    \begin{center}
      \fbox{\parbox[t]{\boxwidth}{\centerline{#2}
          \vspace{2ex}
          \begin{tabular}{lp{\questionwidth}} 
            Given: & #3\\[2pt]
            Question: & #4
          \end{tabular}}}
    \end{center}}}}

\newcommand{\computeproblem}[4]{
  \setlength\boxwidth{\columnwidth}\addtolength\boxwidth{-#1}{
  \setlength\questionwidth{\columnwidth}\addtolength\questionwidth{-#1}\addtolength\questionwidth{-2.5cm}{
    \begin{center}
      \fbox{\parbox[t]{\boxwidth}{\centerline{#2}
          \vspace{2ex}
          \begin{tabular}{lp{\questionwidth}} 
            Given: & #3\\[2pt]
            Task: & #4
          \end{tabular}}}
    \end{center}}}}

\DeclareMathOperator*{\argmax}{arg\,max}

\newcommand{\bestWeightEval}{BestWeightEvaluation}
\newcommand{\bestWeightEnum}{BestWeightEnumeration}

\newcommand{\mx}[1]{\max(#1)}
\newcommand{\mms}{\text{MMS}_\srd}
\newcommand{\enc}[1]{\|#1\|}
\newcommand{\encFunc}{\operatorname{enc}}
 

\begin{document}

\title[Weight Annotation in Information Extraction]{Weight Annotation in Information Extraction{\rsuper*}}
\titlecomment{{\lsuper*}An abridged version of this article has been published in the 2020 International Conference on Database Theory~\cite{DoleschalKMP20}.}

\author[J. Doleschal]{Johannes Doleschal\rsuper{a}}	
\address{University of Bayreuth, Germany; Hasselt University, Belgium}
\email{johannes.doleschal@uni-bayreuth.de}

\author[B. Kimelfeld]{Benny Kimelfeld\rsuper{b}}
\address{Technion, Haifa, Israel}
\email{bennyk@cs.technion.ac.il}

\author[W. Martens]{Wim Martens\rsuper{c}}
\address{University of Bayreuth, Germany}
\email{wim.martens@uni-bayreuth.de}

\author[L. Peterfreund]{Liat Peterfreund\rsuper{d}}
\address{DI-ENS, ENS, CNRS, PSL University, and
	Inria, Paris, France}
\email{liatpf.cs@gmail.com}

\begin{abstract}
The framework of document spanners abstracts the task of information
extraction from text as a function that maps every document (a string) 
into a relation over the document's spans (intervals identified by their
start and end indices). For instance, the regular spanners are the
closure under the Relational Algebra (RA) of the regular expressions
with capture variables, and the expressive power of the regular
spanners is precisely captured by the class of \vset-automata --- a
restricted class of transducers that mark the endpoints of selected spans.

In this work, we embark on the investigation of document spanners that
can annotate extractions with auxiliary information such as
confidence, support, and confidentiality measures. To this end, we
adopt the abstraction of provenance semirings by Green et al., where
tuples of a relation are annotated with the elements of a commutative
semiring, and where the annotation propagates through the positive
RA operators via the semiring operators. Hence, the proposed spanner
extension, referred to as an \emph{annotator}, maps every string into an
annotated relation over the spans.  As a specific instantiation, we
explore weighted \vset-automata that, similarly to weighted automata
and transducers, attach semiring elements to
transitions. We investigate key aspects of expressiveness, such as the
closure under the positive RA, and key aspects of computational
complexity, such as the enumeration of annotated answers and their
ranked enumeration in the case of ordered semirings. For a number of
these problems, fundamental properties of the underlying semiring,
such as positivity, are crucial for establishing tractability.
\end{abstract}

\maketitle

\section{Introduction}
A plethora of paradigms have been developed over the past decades towards the
challenge of extracting structured information from text --- a task generally
referred to as Information Extraction (IE). Common textual sources include
natural language from a variety of sources such as scientific publications,
customer input and social media, as well as machine-generated activity logs.
Instantiations of IE are central components in text analytics and include tasks
such as segmentation, named-entity recognition, relation extraction, and
coreference resolution~\cite{DBLP:journals/ftdb/Sarawagi08}. Rules and rule
systems have consistently been key components in such paradigms, yet their roles
have varied and evolved over time. Systems such as
Xlog~\cite{DBLP:conf/vldb/ShenDNR07} and
SystemT~\cite{DBLP:conf/acl/ChiticariuKLRRV10} use IE rules for materializing
relations inside \e{relational query languages}. Machine-learning classifiers
and probabilistic graphical models (e.g., Conditional Random Fields) use rules
for \e{feature
  generation}~\cite{DBLP:conf/dsmml/LiBC04,DBLP:journals/ftml/SuttonM12}. Rules
serve as \e{weak constraints} (later translated into probabilistic graphical
models) in Markov Logic Networks~\cite{DBLP:conf/aaai/PoonD07} (abbrev. MLNs)
and in the DeepDive system~\cite{DBLP:journals/pvldb/ShinWWSZR15}. Rules are
also used for generating \e{noisy training data} (``labeling functions'') in the
Snorkel system~\cite{DBLP:journals/pvldb/RatnerBEFWR17}.

The framework of \e{document spanners} (\e{spanners} for short) provides a
theoretical basis for investigating the principles of relational rule systems
for IE~\cite{DBLP:journals/jacm/FaginKRV15}. Specifically, a spanner extracts
from a document a relation over text intervals, called \e{spans}, using either
atomic extractors or a relational query on top of the atomic extractors. More
formally, by a \e{document} we refer to a string $\doc$ over a finite alphabet,
a \e{span} of $\doc$ represents a substring of $\doc$ by its start and end
positions, and a \e{spanner} is a function that maps every document $\doc$ into
a relation over the spans of $\doc$. The most studied spanner language is that
of the \e{regular} spanners: atomic extraction is via \e{regex formulas}, which
are regular expressions with capture variables, and relational manipulation is
via the relational algebra: projection, natural join, union, and difference.
Equivalently, the regular spanners are the ones expressible as \e{variable-set
  automata} (\vset-automata for short), which are nondeterministic finite-state
automata that can open and close variables (playing the role of the attributes
of the extracted relation). Interestingly, there has been an independent recent
effort to express artificial neural networks for natural language processing by
means of finite-state automata~\cite{DBLP:conf/icml/WeissGY18,
  DBLP:conf/cdmake/MayrY18, DBLP:conf/iclr/MichalenkoSVBCP19}.

To date, the research on spanners has focused on their expressive
power~\cite{DBLP:journals/jacm/FaginKRV15,DBLP:conf/icdt/PeterfreundCFK19,DBLP:journals/mst/Freydenberger19,SchmidS21},
the computational complexity of their main algorithmic problems~\cite{DBLP:conf/pods/ArenasCJR19,AmarilliBMN19,DBLP:conf/pods/FreydenbergerKP18,FlorenzanoRUVV18},
incompleteness~\cite{DBLP:conf/pods/MaturanaRV18,
  DBLP:conf/pods/PeterfreundFKK19}, and other system aspects such as
cleaning~\cite{DBLP:journals/tods/FaginKRV16} and distributed query
planning~\cite{DoleschalKMNN19}. That research has exclusively adopted
a Boolean approach: \e{a tuple is either extracted or
  not}. Nevertheless, when applied to noisy or fuzzy domains such as
natural language, modern approaches in artificial intelligence adopt a
quantitative approach where each extracted tuple is associated with a
level of confidence that the tuple coincides with the intent.  When
used within an end-to-end IE system, such confidence can be used as a
principled way of tuning the balance between precision and recall. For
instance, in probabilistic IE models (e.g., Conditional Random
Fields), each extraction has an associated probability. In systems of
weak constraints (e.g., MLN), every rule has a numerical weight, and
the confidence in an extraction is an aggregation of the weights of
the invoked rules that lead to the extraction. IE via artificial
neural networks typically involves thresholding over a produced score
or confidence value~\cite{DBLP:journals/tacl/ChiuN16,
  DBLP:journals/tacl/PengPQTY17}. Numerical scores in the extraction
process are also used for quantifying the similarity between
associated substrings, as done with sequence alignment and edit
distance in the analysis of biological sequences such as DNA and
RNA~\cite{torrents2003genome, DBLP:journals/jcb/WangJ94}. 

In this work, we embark on the investigation of spanners that quantify the
extracted tuples. We do so by adopting the concept of \e{annotated relations}
from the framework of \e{provenance semirings} by Green et
al.~\cite{DBLP:conf/pods/GreenKT07}. In essence, every tuple of the database is
annotated with an element of a commutative semiring, and the positive relational
algebra manipulates both the tuples and their annotations by translating
relational operators into semiring operators (e.g., product for natural join and
sum for union). An annotated relation is referred to as a $\srd$-relation, where
$\srd$ is the domain of the semiring. The conceptual extension of the spanner
model is as follows. Instead of a spanner, which is a function that maps every document
$\doc$ into a relation over the spans of $\doc$, we now consider a function
that maps every document $\doc$ into a $\srd$-relation over the spans of $\doc$.
We refer to such a function as a \e{$\srd$-annotator}. Interestingly, as in the
relational case, we can vary the meaning of the annotation by varying the
semiring.
\begin{itemize}
\item We can model \emph{confidence} via the \e{probability} (a.k.a.~\e{inside})
  semiring and the \e{Viterbi} (best derivation)
  semiring~\cite{DBLP:journals/coling/Goodman99}.
\item We can model \emph{support} (i.e., number of derivations) via the \e{counting}
  semiring~\cite{DBLP:journals/coling/Goodman99}.
\item We can model \emph{access control} via the semiring of the \e{confidentiality
    policies}~\cite{DBLP:conf/pods/FosterGT08} (e.g., does the
  extracted tuple require reading top-secret sections? which level
  suffices for the tuple?).
\item We can model the traditional spanners via the \e{Boolean} semiring.
\end{itemize}

As a specific instantiation of $\srd$-annotators, we study the class
of \e{$\srd$-weighted \vset-automata}.
Such automata generalize
\vset-automata in the same manner as weighted automata and weighted
transducers (cf., e.g., the Handbook of Weighted
Automata~\cite{DrosteKV09}): transitions are weighted by semiring
elements, the cost of a run is the product of the weights along the
run, and the weight (annotation) of a tuple is the sum of costs of all
the runs that produce the tuple.
(Again, there has been recent research
that studies the connection between models of artificial neural
networks in natural language processing and weighted
automata~\cite{DBLP:conf/acl/SmithST18}.)

Our investigation answers several fundamental questions about the
class of
$\srd$-weighted \vset-automata:
\begin{enumerate}
\item Is this class closed under the positive relational algebra
  (according to the semantics of provenance
  semirings~\cite{DBLP:conf/pods/GreenKT07})?  
\item What is the complexity of computing the annotation of a tuple?
\item Can we enumerate the annotated tuples as efficiently as we can
  do so for ordinary \vset-automata (i.e., regular document spanners)?
\item In the case of ordered semirings, what is the complexity of enumerating
  the answers in ranked order by decreasing weight?
\end{enumerate}
Our answers are mostly positive and show that $\srd$-weighted \vset-automata
possess appropriate expressivity and tractability properties. As for the last
question, we show that ranked enumeration is intractable and inapproximable for
some of the aforementioned semirings (e.g., the probability and counting
semirings), but tractable for positively ordered and bipotent semirings, such as
the Viterbi semiring.

\subsection*{Comparison to the conference version.}
This article is an extended version of \cite{DoleschalKMP20}, which was published at the 23rd International Conference on Database Theory. In addition to the full proofs, we implemented the following changes: 
\begin{enumerate}
\item We define a more precise cost model (Section~\ref{sec:semiring-encodings}) and use it throughout the entire article for complexity analysis. The cost model in \cite{DoleschalKMP20} used unit-cost for all semiring operations, which is sometimes unrealistic. The model we adopt is more precise in the sense that we encode semiring elements in binary. We define what it means for a semiring to have an \emph{efficient encoding}, which essentially is the case when linear sequences of semiring operations can be performed in polynomial time. Furthermore, we show that the semirings that we are interested in our article indeed have efficient encodings.
\item We extended a result on closure properties (Section~\ref{sec:closureProperties}), namely that joins preserve unambiguity (Lemma~\ref{lem:closureJoin}).
\item We present additional characterizations of recognizable $k$-ary string relations in terms of selectability with $\K$-annotators. We proved a characterization in \cite{DoleschalKMP20} in terms of positive semirings $\K$ (Theorem~\ref{thm:positiveSelect}). We show in Section~\ref{sec:stringsele:beyondpositive} the result can be extended beyond positive semirings. We identify a number of properties on the semiring that are sufficient for the equivalence to hold and show that the equivalence also holds for the \L ukasiewcz semiring, which is not positive.
\end{enumerate}

\subsection*{Organization}
The remainder of this article is organized as follows. We give some required
algebraic background, preliminary definitions and notation in
Section~\ref{sec:prelim}. In Sections~\ref{sec:kannotators}
and~\ref{sec:wvset-automata} we define $\K$-Annotators and weighted
\vset-automata --- a formalism to represent $\K$-Annotators. We study their
fundamental properties in Section~\ref{sec:closureProperties} and various
evaluation and enumeration problems of weighted \vset-automata in
Sections~\ref{sec:eval}, and~\ref{sec:enum}. We conclude in
Section~\ref{sec:conclusion}.

\section{Preliminaries}\label{sec:prelim}

Weight annotators read documents and produce \emph{annotated
  relations}~\cite{DBLP:conf/pods/GreenKT07}, which are relations in which each
tuple is annotated with an element from a commutative semiring. In this section,
we revisit the basic definitions and properties of annotated relations.

\subsection{Algebraic Foundations}\label{sec:algebraicFoundations}
We begin by giving some required background on algebraic structures like monoids
and semirings~\cite{Golan99}.

A \e{commutative monoid} $(\mathbb{M}, \ast, \text{id})$ is an algebraic structure
consisting of a set $\mathbb{M}$, a binary operation $\ast$ and an element
$\text{id} \in \mathbb{M}$, such that:
\begin{enumerate}
\item $\ast$ is associative, i.e., $(a \ast b) \ast c = a \ast (b \ast c)$ for
  all $a,b,c \in \mathbb{M}$, 
\item $\ast$ is commutative, i.e. $a \ast b = b \ast a$ for all $a,b \in
  \mathbb{M}$, and
\item $\text{id}$ is an identity, i.e., $\text{id} \ast a = a$ for all $a \in
  \mathbb{M}$.
\end{enumerate}
We say that a monoid $(\mathbb{M}, \ast, \text{id})$ is \emph{bipotent}, if $a
\ast b \in \{a,b\}$, for every $a,b \in
\mathbb{M}$.

A \e{commutative semiring} $\sr$ is an algebraic structure consisting of a set
$\srd$, containing two elements: the \emph{zero} element $\srzero$ and the
\emph{one} element $\srone$. Furthermore, it is equipped with two binary
operations, namely \emph{addition} $\srplus$ and \emph{multiplication}
$\srtimes$ such that:
\begin{enumerate}
\item $(\srd, \srplus, \srzero)$ and $(\srd, \srtimes, \srone)$ are a
  commutative monoids,
\item multiplication distributes over addition, that is, $(a\srplus b) \srtimes
  c = (a\srtimes c) \srplus (b\srtimes c) $ for all $a,b,c \in \srd$, and
\item $\srzero$ is absorbing for $\srtimes$, that is, $\srzero \srtimes a =
  \srzero$ for all $a \in \srd$.
\end{enumerate}
Furthermore, a semiring is
\emph{positive} if, for all $a,b \in \srd$, the following conditions hold:
\begin{itemize}
\item $\srzero \neq \srone$,
\item if $a \srplus b = \srzero$, then $a = \srzero = b$, and
\item if $a \srtimes b = \srzero$, then $a = \srzero$ or $b = \srzero$.
\end{itemize}
We call a semiring \emph{bipotent}, if its additative monoid is bipotent.

An element $a \in \srd$ is a \emph{zero divisor} if $a \neq \srzero$ and there
is an element $b \in \srd$ with $b \neq \srzero$ and $a \srtimes b = \srzero$.
Furthermore, an element $a \in \srd$ has an \emph{additive inverse}, if there is
an element $b \in \srd$ such that $a \srplus b = \srzero$. In the following, we
will also identify a semiring by its domain $\K$ if the rest is clear from the
context. When we do this for numeric semirings such as $\Q$ and $\N$, we always
assume the usual addition and multiplication.

Given a semiring $\sr$ and a set $\srdp \subseteq \srd$ with $\srzero,\srone \in
\srdp$ such that $\srdp$ is closed under addition and multiplication (that is,
for all $a,b \in \srdp$ it holds that $a \srplus b \in \srdp$ and $a \srtimes b
\in \srdp$) then $(\srdp,\srplus, \srtimes,\srzero,\srone)$ is a
\emph{subsemiring} of $\srd$.

\begin{exa}\label{ex:semirings}
	The following are examples of commutative semirings. It is easy to verify
  that all but the numeric semirings and the {\L}ukasiewcz semiring are
  positive.
  \begin{enumerate}
  \item The \emph{numeric semirings} are $(\Q,+,\cdot,0,1)$, $(\Z,+,\cdot,0,1)$
    and $(\N,+,\cdot,0,1)$.
  \item The \emph{counting semiring} $(\N,+,\cdot,0,1)$.
  \item The \emph{Boolean semiring} $(\B, \vee,\wedge, \false,\true )$ where $\B
    = \{ \false, \true\}$.
  \item The \emph{probability semiring}\footnote{One
      may expect the domain to be $[0,1]$, but this is difficult to obtain while
      maintaining the semiring properties. For instance, defining $a \srplus b$
      as $\min\{a+b,1\}$ would violate distributivity.} $(\Q^+, +, \cdot, 0, 1)$. Rabin~\cite{Rabin63} and
    Segala~\cite{Segala06} define probabilistic automata over this semiring,
    where all transition weights must be between 0 and 1 and the sum of all
    transition weights starting some state, labeled by the same label must be 1.
  \item The \emph{Viterbi semiring} $([0,1], \max, \cdot, 0, 1)$ is used in
    probabilistic parsing~\cite{Droste09}.
  \item The \emph{access control semiring} $\mathbb{A} = (\{P < C < S < T < 0\},
    \min, \max,0,P)$, where $P$ is ``public,'' $C$ is ``confidential,'' $S$ is
    ``secret,'' $T$ is ``top secret,'' and $0$ is ``so secret that nobody can
    access it''~\cite{DBLP:conf/pods/FosterGT08}.
  \item The \emph{tropical semiring} $(\N \cup \{ \infty \},\min, +, \infty, 0)$
    where $\min$ stands for the binary minimum function. This semiring is used
    in optimization problems of networks~\cite{Droste09}.\footnote{In literature
      there are actually multiple different definitions for the tropical
      semiring, e.g., $(\Q \cup \{-\infty \},\max, +, -\infty, 0)$ and $(\Z \cup
      \{ \infty \},\min, +, \infty, 0)$. If not mentioned otherwise, we use the
      tropical semiring as defined here.}
  \item The \emph{\L ukasiewcz semiring}, whose domain is $[0,1]$, with addition
    given by $x \srplus y = \max(x,y)$, with multiplication $x \srtimes y =
    \max(0, x + y - 1)$, zero element $0$, and one element $1$. This semiring is
    used in multivalued logics~\cite{Droste09}.\qed
  \end{enumerate}
\end{exa}

Complexity-wise, we use the RAM model with uniform cost measure and logarithmic
word size~\cite{AhoH74} for our complexity results. That is, we assume that
addition and multiplication of numbers, represented by a logarithmic number of
bits, take constant time. Furthermore, we assume that semiring elements are
encoded in binary. That is, the encoding of a semiring $\srd$, is a function
$\encFunc: \srd \to \{0,1\}^*$, which assigns a binary encoding to every
semiring element. Furthermore, we denote the length\footnote{Note that we do not denote the
  encoding length of semiring elements by $|a|$ to obviate confusions with the
  absolute value function for numbers.} of the encoding of an
element $a \in \srd$ by $\enc{a}$. We discuss semiring encodings into more
detail in Section~\ref{sec:semiring-encodings}.

\subsection{Annotated Relations}
\label{sec:wsr}
We assume infinite and disjoint sets $\dom$ and $\vars$, containing \emph{data
  values} (or simply \emph{values}) and \emph{variables}, respectively. Let
$V\subseteq \vars$ be a finite set of variables. A $V$-\emph{tuple} is a
function $\tup : V \to \vals$ that assigns values to variables in $V$. The
\emph{arity} of $\tup$ is the cardinality $|V|$ of $V$. For a subset $X
\subseteq \vars$, we denote the restriction of $\tup$ to the variables in $X$ by
$\tup \restrict X$. We denote the set of all the $V$-tuples by $\vtup{V}$. We
sometimes leave $V$ implicit when the precise set is not important. For the rest
of this article, we assume that $\sr$ is a commutative semiring. A
$(\srd,\dom)$-\emph{relation $R$ over $V$} is a function $R: \vtup{V}
\rightarrow \srd$ such that its \emph{support}, defined by $\supp{(R)} \df
\{\tup \mid R(\tup) \neq \srzero \}$, is finite. We will also write $\tup \in R$
to abbreviate $\tup \in \supp{(R)}$. The size of a $(\srd,\dom)$-relation $R$ is
the \emph{size} of its support, that is, $|R| \df |\supp{(R)}|$. The
\emph{arity} of a $(\srd,\dom)$-relation over $V$ is $|V|$. When $\dom$ is clear
from the context or irrelevant, we also use \emph{$\srd$-relations} to refer to
$(\srd,\dom)$-relations.

\begin{exa}
  The bottom left table in Figure~\ref{fig:runningExample} shows an example
  $(\srd,\dom)$-relation, where $\K$ is the Viterbi semiring. The variables are
  $x_{\textsf{pers}}$ and $x_{\textsf{loc}}$, so the $V$-tuples are described in
  the first two columns. The third column contains the element in $\K$
  associated to each tuple.\qed
\end{exa}

Green et al.~\cite{DBLP:conf/pods/GreenKT07} defined a set of operators on
$(\srd,\dom)$-relations that naturally correspond to relational algebra
operators and map $\srd$-relations to $\srd$-relations. They define the
algebraic operators\footnote{As in much of the work on semirings in provenance,
  e.g., Green et al.~\cite{DBLP:conf/pods/GreenKT07}, we do not yet consider the
  \emph{difference} operator (which would require additive inverses).}
\emph{union}, \emph{projection}, \emph{natural join}, and \emph{selection} for
all finite sets $V_1, V_2 \subseteq \vars$ and for all $\srd$-relations $R_1$
over $V_1$ and $R_2$ over $V_2$, as follows.
\begin{itemize}
\item \textbf{Union}: If $V_1 = V_2$ then the union $R \df R_1 \cup R_2$ is a
  function $R : \vtup{V_1} \rightarrow \srd$ defined by $R(\tup) \df R_1(\tup)
  \srplus R_2(\tup)$. (Otherwise, the union is not defined.)
\item \textbf{Projection}: For $X\subseteq V_1$, the projection $R\df \pi_X R_1$
  is a function $R: \vtup{X} \rightarrow \srd$ defined by
 	\[
    R(\tup) \df \boplus_{\tup = \tup^\prime \restrict X \text{ and
      }R_1(\tup^\prime)\ne \srzero} R_1( \tup^\prime).
  \]
\item \textbf{Natural Join:} The natural join $R \df R_1 \join R_2 $ is a
  function $R : \vtup{(V_1 \cup V_2)} \rightarrow \srd$ defined by
	$$ 
	R(\tup)  \df R_1(\tup_1) \srtimes R_2(\tup_2)
	$$ 
	where $\tup_1$ and $ \tup_2$ are the restrictions $\tup \restrict V_1$ and
  $\tup \restrict V_2$, respectively.
\item \textbf{Selection:} If $\sel: \vtup{V_1} \rightarrow \{\srzero, \srone\}$
  is a selection predicate that maps each $\vtup{V_1}$ $\tup$ to either
  $\srzero$ or $\srone$ then $R \df \sigma_{\sel}(R_1)$ is a function $R:
  \vtup{V_1} \rightarrow \srd$ defined by
	$$
	R(\tup) \df R_1(\tup) \srtimes {\sel}(\tup).
	$$
\end{itemize}

\begin{propC}[\cite{DBLP:conf/pods/GreenKT07}]
\label{prop:Kalgebra}
The above operators preserve the finiteness of the supports and therefore they
map $\srd$-relations into $\srd$-relations.
\end{propC}
Hence, we obtain an algebra on $\srd$-relations.

\section{K-Annotators}\label{sec:kannotators}

\def\n#1{\textsf{\tiny{#1}}}
\def\s#1{\texttt{\it\em\tt #1}}
\begin{figure}[t]
  \centering\small
  \scalebox{0.7}{
    {\setlength{\tabcolsep}{0.28mm}
      {
        \begin{tabular}{
          cccccccccc
          cccccccccc
          cccccccccc
          cccccccccc
          cccccccccc
          cccccccccc
          cccccccccc
          cccccccccc
          cccccccc
          }
$\s{C}$ & $\s{a}$ & $\s{r}$  & $\s{t}$ & $\s{e}$ & $\s{r}$ & $\s{\blank}$ &
$\s{f}$ & $\s{r}$ & $\s{o}$ & $\s{m}$ &  $\s{\blank}$ &
$\s{P}$ & $\s{l}$ & $\s{a}$  & $\s{i}$ & $\s{n}$ & $\s{s}$ & $\s{,}$ 
          & $\s{\blank}$ &
$\s{G}$ & $\s{e}$ & $\s{o}$  & $\s{r}$ & $\s{g}$ & $\s{i}$ & $\s{a}$ &
          $\s{,}$ &  $\s{\blank}$ &
$\s{W}$ & $\s{a}$ & $\s{s}$  & $\s{h}$ & $\s{i}$ & $\s{n}$ & $\s{g}$ &
          $\s{t}$ & $\s{o}$ & $\s{n}$ & $\s{\blank}$ &
$\s{f}$ & $\s{r}$ & $\s{o}$ & $\s{m}$ &  $\s{\blank}$ &
$\s{W}$ & $\s{e}$ & $\s{s}$  & $\s{t}$ & $\s{m}$ & $\s{o}$ & $\s{r}$ &
          $\s{e}$ & $\s{l}$ & $\s{a}$ & $\s{n}$ & $\s{d}$  & $\s{,}$ & $\s{\blank}$ &
$\s{V}$ & $\s{i}$ & $\s{r}$  & $\s{g}$ & $\s{i}$ & $\s{n}$ & $\s{i}$ & $\s{a}$
          \\\hline
          \n{1} & \n{2} & \n{3} & \n{4} & 
          \n{5} & \n{6} & \n{7} & \n{8} & 
          \n{9} & \n{10}  & \n{11} & \n{12} & 
          \n{13} & \n{14}  & \n{15} & \n{16} & 
          \n{17} & \n{18}  & \n{19} & \n{20} & 
          \n{21} & \n{22}  & \n{23}  & \n{24} & 
          \n{25} & \n{26}  & \n{27} & \n{28} & 
          \n{29} & \n{30}  & \n{31} & \n{32} & 
          \n{33} & \n{34}  & \n{35} & \n{36} &
          \n{37} & \n{38} & 
          \n{39} & \n{40}  & \n{41} & \n{42} & 
          \n{43} & \n{44}  & \n{45} & \n{46} 
          & \n{47} & \n{48} & 
          \n{49} & \n{50}  & \n{51} & \n{52} & 
          \n{53} & \n{54}  & \n{55} & \n{56} &
          \n{57} & \n{58} & 
          \n{59} & \n{60}  & \n{61} & \n{62} & 
          \n{63} & \n{64}  & \n{65} & \n{66} & \n{67}
        \end{tabular}
      }
    }
  }  

  \medskip
  
  \begin{tabular}[t]{ccc}
    \toprule
    $x_{\textsf{pers}}$ & $x_{\textsf{loc}}$ & annotation \\
    \midrule
    Carter & Plains,\blank Georgia & $0.9$ \\
    Washington & Westmoreland,\blank Virginia & $0.9$ \\
    Carter & Georgia,\blank Washington & $0.81$ \\
    Carter & Westmoreland,\blank Virginia & $0.59049$ \\
    \bottomrule
  \end{tabular}
  \qquad
  \begin{tabular}[t]{ccc}
    \toprule
    $x_{\textsf{pers}}$ & $x_{\textsf{loc}}$ & annotation \\
    \midrule
    $\mspan{1}{7}$ & $\mspan{13}{28}$ & $0.9$
    \\
    $\mspan{30}{40}$ & $\mspan{46}{68}$ & $0.9$
    \\
    $\mspan{1}{7}$ & $\mspan{21}{40}$ & $0.81$
    \\
    $\mspan{1}{7}$ & $\mspan{46}{68}$ & $0.59049$
    \\
    \bottomrule
  \end{tabular}  
  \caption{A document (top), a  $(\srd,\dom)$-relation (bottom left), and an
    extracted annotated span relation
    (bottom right). \label{fig:runningExample}}
\end{figure}

  We start by setting the basic terminology. We fix a finite set
$\alphabet$, which is disjoint from $\vars$, that we call the \emph{alphabet}. A
\emph{document} is a finite string over the alphabet $\alphabet$, that is a
finite sequence $\doc = \sigma_1 \cdots \sigma_n$ where $\sigma_i \in \alphabet$
for each $i = 1,\ldots,n$. By $\docs$ we denote the set of all documents. A
\emph{($k$-ary) string relation} is a subset of $\docs^k$ for some $k \in \N$.

A \emph{span} identifies a substring of a document $\doc$ by specifying its
bounding indices, that is, a span of $\doc$ is an expression of the form
$\mspan{i}{j}$ where $1 \le i\le j \le n+1$. By $\doc_{\mspan{i}{j}}$ we denote
the substring $\sigma_i \cdots \sigma_{j-1}$. If $i=j$, it holds that
$\doc_{\mspan{i}{j}}$ is the empty string, which we denote by $\varepsilon$. We
denote by $\spans(\doc)$ the set of all possible spans of a document $\doc$ and
by $\spans$ the set of all possible spans of all possible documents. Since we
will be working with relations over spans, we assume that $\dom$ is such that
$\spans \subseteq \dom$. A \emph{$(\srd,\doc)$-relation} over $V \subseteq
\vars$ is defined analogously to a $(\srd,\dom)$-relation over $V$ but only uses
$V$-tuples with values from $\spans(\doc)$.

\begin{defi}
  A \emph{$\srd$-annotator} (or \emph{annotator} for short), is a function $S$
  that is associated with a finite set $V\subseteq \vars$ of variables and maps
  documents $\doc$ into $(\srd,\doc)$-relations over $V$. We denote $V$ by
  $\vars(S)$. We sometimes also refer to an annotator as an \emph{annotator over
    $\srd$} when we want to emphasize the semiring.
\end{defi}

Notice that $\B$-annotators, i.e., annotators over the Boolean semiring
$(\B,\lor,\land,\false,\true)$ are simply the \emph{document spanners} as
defined by Fagin et al.~\cite{DBLP:journals/jacm/FaginKRV15}.

\begin{exa}
  We provide an example document $\doc$ in Figure~\ref{fig:runningExample}
  (top). The table at the bottom right depicts a possible $(\srd,\doc)$-relation
  obtained by a spanner that extracts $($person, hometown$)$ pairs from $\doc$.
  Notice that for each span $\mspan{i}{j}$ occurring in this table, the string
  $\doc_{\mspan{i}{j}}$ can be found in the table to the left.

  In this na\"ive example, which is just to illustrate the definitions, we used
  the Viterbi semiring and annotated each tuple with $(0.9)^k$, where $k$ is the
  number of words between the spans associated to $x_{\textsf{pers}}$ and
  $x_{\textsf{loc}}$. The annotations can therefore be interpreted as confidence
  scores.\qed
\end{exa}

We now lift the relational algebra operators on $\K$-relations to the level of
$\K$-annotators. For all documents $\doc$ and for all annotators $S_1$ and $S_2$
associated with $V_1$ and $V_2$, respectively, we define the following:
\begin{itemize}
	\item \textbf{Union:} If $V_1= V_2$ then the union $S \df S_1 \cup S_2$ is
    defined by $S(\doc) \df S_1 (\doc)\cup S_2(\doc)$.\footnote{Here, $\cup$
      stands for the union of two $K$-relations as was defined previously. The
      same is valid also for the other operators.} 
	\item \textbf{Projection:} For $X \subseteq V_1$, the projection $S \df \pi_X
    S_1$ is defined by $S(\doc) \df \pi_X S_1(\doc)$. 
	\item \textbf{Natural Join:} The natural join $S \df S_1 \join S_2 $ is
    defined by $S(\doc) \df  S_1(\doc) \join S_2(\doc)$.
  \item \textbf{String selection:} Let $R$ be a $k$-ary string
    relation.\footnote{Recall that a \emph{($k$-ary) string relation} is a
      subset of $\docs^k$.} The string-selection operator
    $\sigma^R$ is parameterized by $k$ variables $x_1,\ldots,x_k$ in $V_1$ and
    can be written as $\sigma^R_{x_1,\ldots,x_k}$. Then the annotator $S \df
    \sigma^R_{x_1,\ldots,x_k}S_1$ is defined as $S(\doc) \df
    \sigma_{\sel}(S_1(\doc))$ where $\sel$ is a selection predicate with
    $\sel(\tup) = \srone$ if $(\doc_{\tup(x_1)},\ldots,\doc_{\tup(x_k)}) \in R$;
    and $\sel(\tup) = \srzero$ otherwise.
\end{itemize}
Due to Proposition~\ref{prop:Kalgebra}, it follows that the above operators form
an algebra on $\srd$-annotators.
 
\section{Weighted Variable-Set Automata}\label{sec:wvset-automata}

In this section, we define the concept of a \e{weighted
  \vset-automaton} as a formalism to represent $\K$-annotators. This
formalism is the natural generalization of
\vset-automata~\cite{DBLP:journals/jacm/FaginKRV15} and weighted
automata~\cite{DrosteKV09}.

Let $V \subseteq \vars$ be a finite set of variables. Furthermore, let
$\varop{V} = \{\vop{v}, \vcl{v} \mid v \in V\}$ be the set of
\emph{variable operations}.\footnote{The operation $\vop{v}$
  represents opening variable $v$ and $\vcl{v}$ represents closing
  variable $v$.} A \emph{weighted variable-set automaton over semiring
  $\K$} (alternatively, a \emph{weighted \vset-automaton} or a
\emph{$\K$-weighted \vset-automaton}) is a tuple
$A \df (\Sigma, V,Q, I, F, \delta)$ where $\Sigma$ is a finite
alphabet; $V\subseteq \vars$ is a finite set of variables; $Q$ is a
finite set of \e{states}; $I : Q \to \srd$ is the \emph{initial weight
  function}; $F: Q \to \srd$ is the \emph{final weight function}; and
$\delta : Q \times (\Sigma \cup \{\varepsilon\} \cup \varop{V} )
\times Q \rightarrow \srd$ is a \emph{($\K$-weighted) transition}
function.

We define the \emph{transitions} of $A$ as the set of triples
$(p,o,q)$ with $\delta(p,o,q) \ne \srzero$. Likewise, the
\emph{initial} (resp., \emph{accepting}) states are those states $q$
with $I(q) \neq \srzero$ (resp., $F(q)\neq \srzero$). A \e{run} $\rn$
of $A$ on a document $\doc \df d_1 \cdots d_n$ is a sequence
\[
 (q_0, i_0) \overset{o_1}{\rightarrow} \cdots \overset{o_{m-2}}{\rightarrow} (q_{m-1}, i_{m-1})
 \overset{o_{m-1}}{\rightarrow}  (q_{m}, i_{m}) 
\]
where
\begin{itemize}
\item $I(q_0) \neq \srzero$ and $F(q_m) \neq \srzero$;
\item $i_0 = 1$, $i_m = n+1$, and $i_j \in \{1,\ldots, n\}$ for each $j \in
  \{1,\ldots,m-1\}$;
\item each $o_j$ is in $\Sigma \cup \{ \varepsilon \} \cup \varop{V}$;
\item $i_{j+1} = i_j$ whenever $o_j \in \{\varepsilon \} \cup \varop{V}$ and
  $i_{j+1} = i_j +1$, otherwise; 
\item $\delta(q_j,o_j,q_{j+1}) \ne \srzero$ for all $j\ge 0$.
\end{itemize}

The \e{weight} of a run is obtained by $\srtimes$-multiplying the weights of its
constituent transitions. Formally, the weight $\weight{\rn}$ of $\rn$ is an
element in $\srd$ given by the expression
\[
  I(q_0) \srtimes  \delta(q_0,o_1,q_1) \srtimes \cdots \srtimes
  \delta(q_{m-1},o_{m-1},q_m) \srtimes F(q_m) .
\]
We call $\rho$ \emph{nonzero} if $\weight{\rn} \neq \srzero$. A run is called
\emph{valid} if, for every variable $v \in V$ the following hold: there is
exactly one index $i$ for which $o_i = \vop{v}$ and exactly one index $j > i$
for which $o_j = \vcl{v}$.

For a nonzero and valid run $\rn$, we define $\tup_{\rn}$ as the $V$-tuple that
maps each variable $v \in V$ to the span $\mspan{i_j}{i_{j^\prime}}$ where
$o_{i_{j}} = \vop{v}$ and $o_{i_{j^\prime}}= \vcl{v}$. We denote the set of all
valid and nonzero runs of $A$ on $\doc$ by
\[
  \Rn{A}{\doc}.
\]
We say that a weighted \vset-automaton $A$ is \emph{functional} if all runs of
$A$ are all valid. Note that this definition is the same as the notion of
functionality of \vset-automata \cite{DBLP:journals/jacm/FaginKRV15}. However,
in contrast to \vset-automata, a $\K$-annotator over an arbitrary commutative
semiring $\srd$ might output the empty relation, even though it has multiple
runs. This can be the case, as the weight of a tuple $\tup$ is the sum of the
weights of all its runs $\rn$ with $\tup_\rn = \tup$ and therefore, the weights
can cancel each other out. This can not happen if $\srd$ is positive (e.g., if
$\srd$ is the Boolean semiring).

Notice that there may be infinitely many nonzero and valid runs of a
weighted \vset-automaton on a given document, due to
\emph{$\varepsilon$-cycles}, which are states $q_1,\ldots,q_k$ such
that $(q_i,\varepsilon,q_{i+1})$ is a transition, also referred to as
an $\varepsilon$-transition, for every $i \in \{1,\ldots,k-1\}$ and
$q_1 = q_k$. Similar to much of the standard literature on weighted
automata (see, e.g., \cite{Esik09}) we assume that weighted
\vset-automata do not have $\varepsilon$-cycles, unless mentioned
otherwise. The reason for this restriction is that automata with such
cycles need $\K$ to be closed under infinite sums for their semantics
to be well-defined.\footnote{The semirings need to fulfill additional
  properties.  Furthermore, distributivity, commutativity and
  associativity must also hold for infinite sums. Such semirings are
  called \emph{complete}~\cite{Mohri2009}.}

As such, if $A$ does not have $\varepsilon$-cycles, then the result of applying
$A$ on a document $\doc$, denoted $\repspnrw{A}(\doc)$, is the
$(\srd,\doc)$-relation $R$ for which
\[
  R(\tup) \df \boplus_{\rn\in \Rn{A}{\doc} \text{ and } \tup = \tup_{\rn} }
 \weight{\rn}.
\]
Note that $\Rn{A}{\doc}$ only contains runs $\rho$ that are valid and nonzero.
If $\tup$ is a $V^\prime$-tuple with $V^\prime \ne V$ then $R(\tup) = \srzero$,
because we only consider valid runs. In addition, $\repspnrw{A}$ is well defined
since every $V$-tuple in the support of $\repspnrw{A}(\doc)$ is a $V$-tuple over
$\spans(\doc)$. 

The \emph{size} of a weighted \vset-automaton $A$ is defined by
\[
  |A| \df |Q|+\sum_{q\in Q}\enc{I(q)}+\sum_{q\in Q}\enc{F(q)}+\sum_{p,q\in
    Q,\; a\in(\Sigma\cup\{\varepsilon\}\cup \varop{V})} \enc{\delta(p,a,q)}\;.
\]

We say that a $\K$-annotator $S$ is \emph{regular} if there exists a weighted
\vset-automaton $A$ such that $S = \repspnrw{A}$. Note that this is an equality
between functions. Furthermore, we say that two weighted \vset-automata $A$ and
$A^\prime$ are equivalent if they define the same $\K$-annotator, that is
$\repspnrw{A} = \repspnrw{A^\prime}$, which is the case if $\repspnrw{A}(\doc) =
\repspnrw{A^\prime}(\doc)$ for every $\doc \in \docs$. Similar to
our terminology on $\B$-annotators, we use the term \emph{$\B$-weighted
  \vset-automata} to refer to the ``classical'' \vset-automata of Fagin et
al.~\cite{DBLP:journals/jacm/FaginKRV15}, which are indeed weighted
\vset-automata over the Boolean semiring.

We say that a $\K$-weighted \vset-automaton $A$ is \emph{unambiguous} if, for
every document $\doc$ and every tuple $\tup \in \repspnrk{A}{\K}(\doc)$, there
exists exactly one valid run $\rho$ of $A$ on $\doc$, such that $\tup =
\tup_\rho$, and there is no valid run for tuples $\tup \notin
\repspnrk{A}{\K}(\doc)$.
Note that, over some semirings, the class
of unambiguous weighted \vset-automata is strictly contained in the class of
weighted \vset-automata, as shown in the following proposition. However, over
the Boolean semiring, every $\B$-weighted automaton can be determinized (c.f.
Doleschal et al.~\cite[Proposition 4.4]{DoleschalKMNN19}\footnote{We note that
  a notion of determinism was also introduced by Maturana et
  al.~\cite{DBLP:conf/pods/MaturanaRV18}, but the one in \cite{DoleschalKMNN19} is stronger, while allowing the same expressiveness. An important difference between the two is that the containment problem for deterministic \vset-automata is PSPACE-complete when using the definition of~\cite{DBLP:conf/pods/MaturanaRV18}, whereas it is NL-complete when using the definition of~\cite{DoleschalKMNN19}. A discussion can be found in
  Doleschal et al.~\cite{DoleschalKMNN19}.}). Therefore there is also an
unambiguous $\B$-weighted automaton $A_u$ which is equivalent to $A$, as every
deterministic $\B$-weighted automaton is also unambiguous.

\begin{prop}\label{prop:equivalenceUndec}
  Let $\srd = (\Z \cup \{ \infty \},\min, +, \infty, 0)$ be the tropical
  semiring. There is a $\srd$-weighted \vset-automaton $A$ such that there is no
  $\srd$-weighted unambiguous \vset-automaton $A'$ which is equivalent to $A$.
\end{prop}
\begin{proof}
  Weighted automata can be seen as weighted \vset-automata over the empty set of
  variables. Thus, the statement follows directly from Kirsten~\cite[Proposition
  3.2]{Kirsten08} who showed that there is a $\srd$-weighted automaton $A$ such
  that there is no equivalent unambiguous $\srd$-weighted automaton
  $A'$.\footnote{Actually, Kirsten~\cite[Proposition 3.2]{Kirsten08} showed an
    even stronger result. He showed that, the result still holds if $A$ is a
    polynomially ambiguous weighed automaton, i.e., a weighted automaton for
    which the number of accepting runs of a word of length $n$ is bound by a
    fixed polynomial $p(n)$.}
\end{proof}

\begin{figure}
  \resizebox{\linewidth}{!}{
    \begin{tikzpicture}[>=latex, ->]
      \def\d{2.5}
      \node[state,initial] (q0) at (0,0){$q_0$};
      \node[state] (q1) at ($(q0)+(0,\d)$) {$q_1$};
      \node[state] (q2) at ($(q0)+(\d,0)$) {$q_2$};
      \node[state] (q3) at ($(q2)+(\d,0)$) {$q_3$};
      \node[state] (q4) at ($(q3)+(\d,0)$) {$q_4$};
      \node[state] (q5) at ($(q4)+(\d,0)$) {$q_5$};
      \node[state] (q6) at ($(q5)+(0,\d)$) {$q_6$};
      \node[state] (q7) at ($(q5)+(\d,0)$) {$q_7$};
      \node[state] (q8) at ($(q7)+(\d,0)$) {$q_8$};
      \node[state, accepting ] (q9) at ($(q8)+(\d,0)$) {$q_9$};
      \node[state, accepting] (q10) at ($(q9)+(0,\d)$) {$q_{10}$};
      
      \draw (q0) edge [bend left] node[left] {$\Sigma^\prime;1$}(q1);
      \draw (q1) edge [bend left] node[right] {$\blank;1$}(q0);
      \draw (q1) edge [loop right] node[right] {$\Sigma^\prime;1$}(q1);
      \draw (q0) edge node[below] {$\vop{x_{\textsf{pers}}};1$}(q2);
      \draw (q2) edge[decorate, decoration={snake, amplitude=.5mm, segment length=2mm, post length=1mm}] node[above] {Pers$;1$}(q3);
      \draw (q3) edge node[below] {$\vcl{x_{\textsf{pers}}};1$}(q4);
      \draw (q4) edge node[below] {$\blank;1$}(q5);
      \draw (q5) edge [bend left] node[left] {$\Sigma^\prime;1$}(q6);
      \draw (q6) edge [bend left] node[right] {$\blank;0.9$}(q5);
      \draw (q6) edge [loop right] node[right] {$\Sigma^\prime;1$}(q6);
      \draw (q5) edge node[below] {$\vop{x_{\textsf{loc}}};1$}(q7);
      \draw (q7) edge[decorate, decoration={snake, amplitude=.5mm, segment length=2mm, post length=1mm}] node[above] {Loc$;1$}(q8);
      \draw (q8) edge node[below] {$\vcl{x_{\textsf{loc}}};1$}(q9);
      \draw (q9) edge node[right] {$\blank;1$}(q10);
      \draw (q10) edge [loop left] node[left] {$\Sigma;1$}(q10);
    \end{tikzpicture}}
  \caption{An example weighted \vset-automaton over the Viterbi semiring with
    initial state $q_0$ (with weight $1$), two final states $q_{9},q_{10}$ (both
    with weight $1$), and alphabet $\Sigma^\prime = \Sigma \setminus
    \{\blank\}$. Pers and Loc are sub-automata matching person and location
    names respectively. All edges, including the edges of the sub-automata, have
    the weight $1$ besides the transition from $q_6$ to $q_5$ with weight $0.9$.
  }
    \label{fig:persLocSpanner} 
\end{figure}
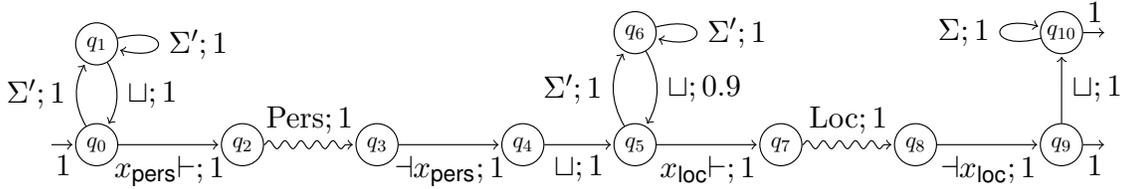

\begin{exa}\label{ex:weightedVset}
  Figure~\ref{fig:persLocSpanner} shows an example weighted \vset-automaton over
  the Viterbi semiring, which is intended to extract (person, hometown)-tuples
  from a document. Here, ``Pers'' and ``Loc'' should be interpreted as
  sub-automata that test whether a string could be a person name or a location.
  (Such automata can be compiled from publicly available regular
  expressions\footnote{For example, \url{https://regexlib.com/}.} and from
  deterministic rules and dictionaries as illustrated in
  SystemT~\cite{DBLP:conf/acl/ChiticariuKLRRV10}.)

  The relation extracted by this automaton from the document in
  Figure~\ref{fig:runningExample} is exactly the annotated span relation of the
  same figure. The weight of a tuple $\tup$ depends on the number of spaces
  occurring between the span captured by $x_\textsf{pers}$ and the span captured
  by $x_\textsf{loc}$. More specifically the automaton assigns the weight
  $(0.9)^k$ to each tuple, where $k$ is the number of words between the two
  variables. \qed
\end{exa}

As we see next, checking equivalence of weighted \vset-automata is undecidable
in general. 
\begin{prop}\label{prop:equivUndec}
	Given two weighted \vset-automata $A_1$ and $A_2$ over the tropical semiring,
  it is undecidable to test whether $\repspnrw{A} = \repspnrw{A'}$.
\end{prop}
\begin{proof}
  Follows directly from undecidability of the containment problem of weighted
  automata over the tropical semiring.(c.f. Krob~\cite[Corollary
  4.3]{Krob94}\footnote{The proof by Krob is quite algebraic. See Almagor et
    al.~\cite[Theorem 4]{ShaullUO11} for an automata theoretic proof.})
\end{proof}

\subsection{Connection to Datalog over Annotated Relations}\label{sec:datalog}
The semantics of $\repspnrw{A}(\doc)$ is similar in spirit to the
semantics of Datalog over annotated relations, studied by Deutch et
al.~\cite{DBLP:conf/icdt/DeutchMRT14}, when we view the runs of
$\rn\in\Rn{A}{\doc} $ as the \e{derivations} of $\tup$: we take the
product of the items that participate in each derivation, and sum up
these products over all derivations. In fact, there is a simple
translation of a weighted vset-automaton into a Datalog program over
annotated relations, similarly to the way Peterfreund et
al.~\cite{DBLP:conf/icdt/PeterfreundCFK19} represent ordinary spanners
via Datalog. Roughly speaking, in this translation the document $\doc$
is conventionally represented via the relations over the positions
such as $\mathsf{O}_\sigma(i)$ and $\mathsf{Successor}(i,j)$ that
store the positions with the symbol $\sigma$ and successor
relationship between positions, respectively (in addition to the
relations $\mathsf{First}(i)$ and $\mathsf{Last}(i)$ that represent
the first and last positions, respectively). Each tuple in these
relations is annotated by $\srone$. In addition, the transitions are
represented by relations of the form $T_\sigma (q,q)$ annotated with
the weight of the corresponding transition. (For simplicity, we ignore
the initial and final weight functions.) The runs are derived by
simple path rules such as
\begin{align*}
  \mathsf{Path}(x,y,q,q')&\leftarrow T_\sigma (q,q'), 
  \mathsf{Successor}(x,y), \mathsf{O}_\sigma(y)\\
\mathsf{Path}(x,y,q,q') &\leftarrow \mathsf{Path}(x,z,q,q''),
                          \mathsf{Path}(x,z,q'',q')
          \end{align*}
 where $\mathsf{Path}(x,y,q,q')$ states that it is possible to reach
 $q'$ from $q$ when starting in $x$ and ending with $y$. 

 It is easy to show that a translation such as the above preserves the
 provenance as defined by Deutch et
 al.~\cite{DBLP:conf/icdt/DeutchMRT14} (i.e., the sum of products of
 tuples in the derivations).  We further discuss this translation, as
 well the general relationship between our work and that of Deutch et
 al.~\cite{DBLP:conf/icdt/DeutchMRT14}, in
 Section~\ref{sec:conclusion}.

\section{Semiring Encodings}\label{sec:semiring-encodings}
In order to state complexity results, we need to make some assumptions
about the representation and computation of the semiring operations.
That is, as mentioned in Section~\ref{sec:algebraicFoundations}, we
assume that semiring elements are encoded in binary, i.e., there is a
function $\encFunc: \srd \to \{0,1\}^*$, which assigns a binary
encoding to every semiring element. We write the length of the
encoding of an element $a \in \srd$ as $\enc{a}$.

Throughout this article, we sometimes encode computations into matrix
multiplications. To this end, we define a \emph{matrix multiplication system
  $\mms$ of dimension $n \in \N$} as a triple $\mms \df (I,M,F)$, where $I,F \in \srd^n$ are $n$-dimensional
vectors over $\srd$ and $M \in \srd^{n\times n}$ is an $n\times n$ matrix. We
define the \emph{size}\ of a matrix multiplication system as its dimension plus
the sum of the encoding lengths of all semiring elements in the system. That is,
\[
  |\mms| = n + \sum\limits_{a \in I} \enc{a} + \sum\limits_{a \in M} \enc{a} +
  \sum\limits_{a \in F} \enc{a}\;.
\]

For an $n \times n$ matrix $X \in \srd^{n\times n}$ (resp., a vector $X\in
\srd^n$), we define $\mx{X}$ to be the maximum of the dimension of $X$ and the
largest encoding length of a semiring element in $X$, that is,
\[
  \mx{X} \df \max(n,\max_{x \in X} \enc{x})\;.
\]
Furthermore, for a matrix multiplication system $\mms = (I,M,F)$, we define
\[
  \mx{\mms} = \max(\mx{I},\mx{M},\mx{F})\;.
\]

Let $F^T$ be the transpose of vector $F$. By $I \times M$ we denote the matrix
multiplication of $I$ and $M$. We define \emph{efficient semiring encodings} as
follows.

\begin{defi}\label{def:efficientSemiring}
   Let $\sr$ be a semiring. The encoding of $\srd$ is \emph{efficient} if, for every matrix multiplication system $\mms$ and every natural number $k$, the encodings of the semiring
  elements
  \[
    w_i \df I \times M^i \times F^T\;,
  \]
  for all $0 \leq i \leq k$, and 
  \[
    w \df \boplus_{1 \leq i \leq k} w_i
  \]
  can be computed in time polynomial in $|\mms| \cdot k$.
\end{defi}

Throughout this section, whenever we give complexity bounds, we assume that an
efficient encoding of the semiring is used. As we show now, the standard
encodings of most of the semirings in Example~\ref{ex:semirings} are efficient.

\begin{prop}\label{prop:noBlowupSemiring}
  Let $\sr$ be a semiring. Then the encoding of $\srd$ is efficient if, for all
  semiring elements $a,b \in \srd$, the encodings of $a \srplus b$ and $a \srtimes b$ can be
  computed in time polynomial in $\enc{a}+\enc{b}$ and
  \begin{align*}
    \enc{a\srplus b} &\leq \max(\enc{a},\enc{b})+1,\text{and}\\
    \enc{a\srtimes b} &\leq \enc{a}+\enc{b}\;.
  \end{align*}
\end{prop}
\begin{proof}
  Let $\mms = (I,M,F)$ be a matrix multiplication system of dimension $n$ and $k
  \in \N$ be a natural number. Let $w_1,\ldots,w_k$ and $w$ be as in Definition~\ref{def:efficientSemiring}. 

  We observe that the computation of $w$ requires a polynomial number of
  additions and multiplications. However, as the encoding of the semiring
  elements that are used for the computation might become large, this does
  not immediately imply that $w$ can be computed in time polynomial in
  $|\mms|\cdot k$.
  
  We therefore show, for every $1 \leq i \leq k$, that the semiring elements that are required for the computation of $w_i$ have an encoding of size
  at most polynomial in $|\mms|\cdot k$. Due to the assumption that $\enc{a
    \srplus b} \leq \max(\enc{a},\enc{b})+1$, this immediately implies that
  $\enc{w}$ is polynomial in $|\mms|\cdot k$, which concludes the proof.
  
  Recall that $\mx{M}$ is the maximum of the dimension $n$ of $M$ and the largest
  representation size of any element in $M$. We
  begin by showing that $\mx{I\times M} \leq \mx{I} + \mx{M} + n$ for all
  vectors $I \in \srd^n$ and matrices $M \in \srd^{n\times n}$. Let $x$ be an
  element of $I \times M$. Per definition of matrix multiplication, $x$ is the
  sum of $n$ elements $x_1,\ldots, x_n$, each of which is the product of an
  element from $I$ and an element from $M$. Thus
  \[
    \enc{x_i} \leq \mx{I} + \mx{M}\;
  \]
  and, therefore,
  \[
    \enc{x} \leq \mx{I} + \mx{M} + n\;.
  \]
  We conclude that $\mx{I\times M} \leq \mx{I} + \mx{M} + n$ for all
  vectors $I \in \srd^n$ and matrices $M \in \srd^{n\times n}$.

  We now show by induction that, for all $i \in \N$, it holds that
  \[
    \mx{I\times M^i \times F^T} \leq \mx{I} + i\cdot (\mx{M}+n) + \mx{F} + n\;,
  \]
  for all vectors $I,F \in \srd^n$ and all matrices $M \in \srd^{n\times n}$. Since $\mx{I} + \mx{M} + \mx{F} + n \leq |\mms|$, this implies
  that $\enc{w_i}$ is polynomial in $|\mms|\cdot k$, for all $0 \leq i \leq k$.

  For the base case, we observe that $w_0 = I \times F$ is the sum of $n$
  elements, each of which has size at most $\mx{I} + \mx{F}$. As desired, we
  therefore have that 
  \[
    \mx{I\times F} = \enc{w_0} \leq \mx{I} + \mx{F} + n\;.
  \]

  For the inductive step, assume there is an $i \in \N$ such that
  \[
    \mx{I\times M^i \times F^T} \leq \mx{I} + i\cdot (\mx{M}+n) + \mx{F} + n
  \]
  for all vectors $I,F \in \srd^n$ and all matrices $M \in \srd^{n\times n}$.
  With $I' \df I \times M$, we have that
  \[
    \mx{I'} = \mx{I\times M} \leq \mx{I} + \mx{M} + n\;
  \]
  and, therefore,
  \begin{align*}
    \mx{I\times M^{i+1} \times F^T}
    & = \mx{I'\times M^i \times F^T} \\
    & \leq \mx{I'} + i\cdot (\mx{M}+n) + \mx{F} + n \\
    & \leq \mx{I} + \mx{M} + n + i\cdot (\mx{M}+n) + \mx{F} + n \\
    & = \mx{I} + (i+1)\cdot (\mx{M}+n) + \mx{F} + n\;.
  \end{align*}
  This concludes the proof.
\end{proof}

Note that all semirings over a finite domain have an efficient encoding, as each
semiring element can be encoded with constant size and all operations can be
carried out in constant time via a constant size lookup table.
\begin{cor}
  Each semiring $\sr$ where $\srd$ is finite has an efficient encoding.
\end{cor}

We observe that, for many semirings, the standard encodings satisfy the
conditions of Proposition~\ref{prop:noBlowupSemiring}. Examples are the numeric
semiring $(\Z,+,\cdot,0,1)$, the counting semiring, the Boolean semiring, the
Viterbi semiring (over the rationals $\Q$), the access control semiring,
and the tropical semirings. However, for some semirings, standard encodings of
the semiring elements do not satisfy the conditions of
Proposition~\ref{prop:noBlowupSemiring}. For example, consider the numeric
semiring $(\Q,+,\cdot,0,1)$ and the encoding, where every semiring element $a =
\frac{n}{d}$ is encoded by its numerator $n \in \Z$ and its denominator $d
\in \N$. The problem is that the sum of two rational numbers
$\frac{a}{b},\frac{c}{d}$ is given by $x = \frac{a}{b} + \frac{c}{d} =
\frac{a\cdot d + b\cdot c}{b\cdot d}$ and therefore the size of the encoding of
$x$ is $\enc{x} \leq \enc{a\cdot d}+\enc{b\cdot c}+\enc{c\cdot d}$ which, in
general, does not satisfy the condition that $\enc{\frac{a}{b}+\frac{c}{d}} \leq
\max(\enc{\frac{a}{b}},\enc{\frac{c}{d}})+1$. Even though this only increases
the size of the representation by a small margin, we need some further investigation to conclude that this encoding is efficient.

\begin{prop}\label{prop:numericEfficientEncoding}
  The numeric semiring $(\Q,+,\cdot,0,1)$ has an efficient encoding.
\end{prop}
\begin{proof}
  Let $(\Q,+,\cdot,0,1)$ be the numeric semiring. We assume that every semiring
  element $x = \frac{a}{b}$ is encoded by its numerator $a \in \Z$ and its
  denominator $b \in \N$. Let all numerators and denominators be encoded in
  binary, where two's complement encoding is used for the numerators. We observe
  that Proposition~\ref{prop:noBlowupSemiring} holds for both encodings.
  Furthermore, the encoding of the denominators is monotone, that is, for every
  $x,y \in \N$ it holds that $\enc{x} \leq \enc{y}$ if $x \leq y$.

  For a matrix multiplication system $\mms = (I,M,F)$, let $D$ be the set of
  all denominators of the rationals in $I$, $F$, and $M$. We will compute the least
  common multiple $d_{\text{lcm}}$ of all denominators in $D$ and expand the
  representations of all numbers to the denominator $d_{\text{lcm}}$. Observe
  that all denominators $d \in D$ are natural numbers. Therefore,
  \[
    \enc{d_{\text{lcm}}} \overset{(1)}{\leq}\Big\|\prod\limits_{d \in D}
    d\;\Big\| \overset{(2)}{\leq} \sum\limits_{d \in D} \enc{d}
    \overset{(3)}{\leq} |\mms|\;,
  \]
  where inequality $(1)$ follows from $d_{\text{lcm}} \leq \prod\limits_{d \in D}d$ and the
  monotonicity of the encoding of the denominators, inequality $(2)$ follows from $\enc{x
    \cdot y} \leq \enc{x} + \enc{y}$, and inequality $(3)$ follows from the definition
  of $|\mms|$. Therefore, $\enc{d_{\text{lcm}}} \leq |\mms|$ is polynomial in
  $|\mms|$. Furthermore, the computation of $d_{\text{lcm}}$ as well as the
  expansion can be done in polynomial time.\footnote{The least common multiple
    can be computed using the Eucledian algorithm and the expansion of $x =
    \frac{a}{b}$ by multiplying the numerator $a$ by
    $\frac{b}{d_{\text{lcm}}}$.} We therefore assume \mbox{w.l.o.g.} that all
  rationals in $I,F$, and $M$ have the denominator $d_{\text{lcm}}$.

  Let $I_\Z, F_\Z \in \Z^n$ and $M_\Z \in \Z^{n\times n}$ be the
  vectors $I,F$ and the matrix $M$ where all numbers are replaced by their
  numerator. For all $1 \leq i \leq k$, we define
  \[
    w_{\Z, i} \df I_\Z \times M_\Z^i \times F_\Z^T\;.
  \]
  We recall that, due to Proposition~\ref{prop:noBlowupSemiring}, $w_{\Z, i}$
  can be computed in time polynomial in $|\mms|\cdot k$. Per assumption that all
  rationals in $I$, $F$, and $M$ have the denominator $d_{\text{lcm}}$, we have that
  $w_i = \frac{w_{\Z, i}}{d_{\text{lcm}}^{i+2}}$. Furthermore, the
  denominator can also be computed in time polynomial in $|\mms|$, as
  $\enc{d_{\text{lcm}}}$ is polynomial in $|\mms|$ and $\enc{x\cdot
    y}\leq \enc{x}+\enc{y}$ for the encoding of natural numbers. Thus, for all
  $i \leq k$, the encodings of the $w_i$ can be computed in time polynomial in $|\mms|\cdot k$.
  Furthermore, $w$ can be computed in time polynomial in $|\mms|\cdot k$ by
  first expanding all $w_i$ to the denominator $d_{\text{lcm}}^{k+2}$ and
  summing up the expanded fractions. This concludes the proof.
\end{proof}
 
\section{Fundamental Properties}\label{sec:closureProperties}
We now study fundamental properties of annotators. Specifically, we show that
regular annotators are closed under union, projection, and join. Furthermore,
annotators over a semiring $\K$ behave the same as document spanners with
respect to string selection if $\K$ is positive or $\srplus$ is
bipotent\footnote{Recall, $\srplus$ is bipotent, if $a \srplus b \in \{a,b\}$,
  for every $a,b \in \K$.} and for every $a,b \in \srd$, $a \srtimes b = \srone$
implies that $a = b = \srone$.

\subsection{Epsilon Elimination}
We begin the section by showing that every regular $\K$-annotator can be
transformed into an equivalent functional regular $\K$-annotator without
$\varepsilon$-transitions.

\begin{prop}\label{prop:epsilonRemoval}
  For every weighted \vset-automaton $A$ there is an equivalent weighted
  \vset-automaton $A^\prime$ that has no $\varepsilon$-transitions. This
  automaton $A^\prime$ can be constructed from $A$ in polynomial time.
  Furthermore, $A$ is functional if and only if $A^\prime$ is functional.
\end{prop}
\begin{proof}
  We use a result by Mohri~\cite[Theorem 7.1]{Mohri2009} who showed that, given
  a weighted automaton, one can construct an equivalent weighted automaton
  without epsilon transitions.

  More precisely, let $A = (\Sigma,V,Q,I,F,\delta)$ be a weighted
  \vset-automaton. Notice that $A$ can also be seen as an ordinary weighted
  finite state automaton $B = (\Sigma \cup \varop{V}, Q,I,F,\delta)$. In this
  automaton, one can remove epsilon transitions by using Mohri's epsilon removal
  algorithm~\cite[Theorem 7.1]{Mohri2009}. The resulting
  $\varepsilon$-transition free automaton $B^\prime = (\Sigma \cup
  \varop{V},Q^\prime,I^\prime,F^\prime,\delta^\prime)$ accepts the same strings
  as $B$. Therefore, interpreting $B^\prime$ as an weighted \vset-automaton
  $A^\prime = (\Sigma, V,Q^\prime,I^\prime,F^\prime,\delta^\prime)$ we have that
  $\repspnrw{A} = \repspnrw{A^\prime}$ and $A^\prime$ is functional if and only
  if $A$ is functional.

  Concerning complexity, Mohri shows that this algorithm runs in polynomial
  time, assuming that weighted-$\varepsilon$-closures can be computed in
  polynomial time. However, in our setting this is obvious as we do not allow
  $\varepsilon$-cycles. Therefore, the weight of an element of an
  $\varepsilon$-closure can be computed by at most $n$ matrix multiplications,
  where $n$ is the number of states\footnote{As such, the construction
    also works in a slightly more general setting than ours, where the semiring
    is \emph{complete} (closed under taking infinite sums, associativity,
    commutativity, and distributivity apply for countable sums) and weights of
    $\varepsilon$-closures can be computed in polynomial time.} in $A$. Per assumption
  that $\srd$ has an efficient encoding, these matrix multiplications can be
  computed in polynomial time.
\end{proof}

\subsection{Functionality}
Non-functional \vset-automata are inconvenient to work with, since some of their
nonzero runs are not valid and therefore do not contribute to the weight of a
tuple. It is therefore desirable to be able to automatically convert weighted
\vset-automata into functional weighted \vset-automata.

\begin{prop}\label{prop:functionalNormalform}
  Let $A$ be a weighted \vset-automaton. Then there is a functional weighted
  \vset-automaton $A_\text{fun}$ that is equivalent to $A$. If $A$ has $n$
  states and uses $k$ variables, then $A_\text{fun}$ can be constructed in time
  polynomial in $n$ and exponential in $k$.
\end{prop}
\begin{proof}
  The proof follows the idea of a similar result by
  Freydenberger~\cite[Proposition 3.9]{DBLP:journals/mst/Freydenberger19} for
  unweighted \vset-automata. Like Freydenberger, we associate each
  state in $A_{\text{fun}}$ with a function $s: V \to \{w,o,c\}$, where $s(x)$
  represents the following:
  \begin{itemize}
    \item $w$ stands for ``waiting,'' meaning $\vop{x}$ has not been read,
    \item $o$ stands for ``open,'' meaning $\vop{x}$ has been read, but not $\vcl{x}$,
    \item $c$ stands for ``closed,'' meaning $\vop{x}$ and $\vcl{x}$ have been read.
  \end{itemize}
  
  Let $S$ be the set of all such functions. Observe that $|S| = 3^{|V|}$. We now
  define $A_{\text{fun}} \df (\Sigma, V, Q_{\text{fun}}, I_{\text{fun}}, F_{\text{fun}},
  \delta_{\text{fun}})$ as follows:
  \begin{align*}
    Q_{\text{fun}} &\df Q \times S\\
    I_{\text{fun}}(p,s) &\df
    \begin{cases}
        I(p) & \text{where } s(x) = w \text{ for all } x \in V\\
        \srzero & \text{otherwise}
    \end{cases} \\
    F_{\text{fun}}(p,s) &\df
      \begin{cases}
        F(p) & \text{where } s(x) = c \text{ for all } x \in V\\
        \srzero & \text{otherwise}
      \end{cases} \\
  \end{align*}
  Furthermore, for all $(p,s) \in Q_{\text{fun}}$ and $x \in V$ we define
  \[
    \begin{array}{r@{\ }ll}
      \delta_{\text{fun}}((p,s), a, (q,s)) & = \delta(p,a,q) & \text{for } a \in (\Sigma \cup \{\varepsilon\}),\\
      \delta_{\text{fun}}((p,s), \vop{x}, (q,s^x_o)) & = \delta(p,\vop{x},q) & \text{if } s(x) = w,\\
      \delta_{\text{fun}}((p,s), \vcl{x}, (q,s^x_c)) & = \delta(p,\vcl{x},q) & \text{if } s(x) = o,\\
      \delta_{\text{fun}}((p,s), a, (q,t)) & = \srzero & \text{otherwise},\\
    \end{array}
  \]

  \noindent where $s^x_o$ is defined by $s^x_o(x) \df o$, and $s^x_o(y) \df
  s(y)$ for all $x \neq y$, and $s^x_c$ is defined by $s^x_c(x) \df c$ and
  $s^x_c(y) \df s(y)$ for all $x \neq y$.

  Functionality follows analogously to Freydenberger~\cite[Proposition
  3.9]{DBLP:journals/mst/Freydenberger19}. It remains to show equivalence, i.e.,
  that for every document $\doc \in \docs$ it holds that $\repspnrw{A}(\doc) =
  \repspnrw{A_{\text{fun}}}(\doc)$. Observe that there is a one to one
  correspondence between valid nonzero runs $\rn \in \Rn{A}{\doc}$ and valid
  nonzero runs $\rn_{\text{fun}} \in \Rn{A_{\text{fun}}}{\doc}$ with
  $\weight{\rn} = \weight{\rn_{\text{fun}}}$. Therefore, $\repspnrw{A}(\doc) =
  \repspnrw{A_{\text{fun}}}(\doc)$ must also hold.
\end{proof}

The exponential blow-up in Proposition~\ref{prop:functionalNormalform} cannot be
avoided, since it already occurs for \vset-automata over the Boolean
semiring.\footnote{Freydenberger~\cite[Proposition
  3.9]{DBLP:journals/mst/Freydenberger19} showed that there is a class of
  \vset-automata $\{A_k \mid k \in \N\}$ (over $\B$), each with one state and
  $k$ variables, such that every functional \vset-automaton equivalent to $A_k$
  has at least $3^k$ states.} Functionality of \vset-automata can be checked
efficiently, as we have the following result.
\begin{prop}\label{prop:functional}
  Given a $\K$-weighted \vset-automaton $A$ with $m$ transitions and $k$
  variables, it can be decided whether $A$ is functional in time $O(km)$.
  Furthermore, $A$ is functional if and only if it is functional when
  interpreted as $\B$-weighted document spanner.
\end{prop}
\begin{proof}
  Per definition, a weighted \vset-automaton is functional if all runs are
  valid. Furthermore, a run $\rn$ is valid if for every variable $v \in V$ there
  is exactly one index $i$ for which $o_i = \vop{v}$ and exactly one index $j >
  i$ for which $o_j = \vcl{v}$.

  Observe that this definition only depends on the labels of the run and not on
  the semiring of the automaton. Therefore, a $\K$-weighted \vset-automaton $A$
  is functional if and only if $A$ is functional when interpreted as an
  $\B$-weighted \vset-automaton $A^\B$. More formally, let $A^\B$ be the
  $\B$-weighted \vset-automaton obtained by replacing nonzero weights with
  $\true$, sum by $\lor$ and multiplication by $\land$. The result now follows
  directly from Freydenberger~\cite[Lemma
  3.5]{DBLP:journals/mst/Freydenberger19}, who showed that it can be verified in
  $O(km)$ whether a \vset-automaton is functional.
\end{proof}

 \subsection{Closure Under Join, Union, and Projection}

We will obtain the following result.
\begin{thm}\label{theo:closed-algebra}
  Regular annotators are closed under finite union, projection, and finite
  natural join. Furthermore, if the annotators are given as functional weighted
  \vset-automata, the construction for a single union, projection, and join can
  be done in polynomial time. Furthermore, the constructions preserve
  functionality.
\end{thm}
The theorem follows immediately from Lemmata~\ref{lem:closureUnion},
\ref{lem:closureProjection}, and \ref{lem:closureJoin}. Whereas the
constructions for union and projection are fairly standard, the case of join
needs some care in the case that the two automata $A_1$ and $A_2$ process
variable operations in different orders.\footnote{More formally, if $A_1$
  processes $\vop{x}\vop{y} a \vcl{y}\vcl{x}$ and $A_2$ processes
  $\vop{y}\vop{x}a\vcl{x}\vcl{y}$, then these two different sequences produce
  different encodings of the same tuple. This has to be considered by the
  automata construction.}

\begin{lem}\label{lem:closureUnion}
	Given two $\K$-weighted \vset-automata $A_1$ and $A_2$ with $V_1 = V_2$, one can
  construct a weighted \vset-automaton $A$ in linear time, such that 
  $\repspnrw{A} = \repspnrw{A_1} \cup \repspnrw{A_2}$.
\end{lem}
\begin{proof}
  This lemma follows by the standard construction for the union of two weighted
  automata.
  
  Let $A_1 \df (\Sigma, V,Q_1,I_1,F_1,\delta_1)$ and $A_2 \df (\Sigma,
  V,Q_2,I_2,F_2,\delta_2)$, such that $Q_1 \cap Q_2 = \emptyset$. We construct
  an automaton $A \df (\Sigma, V,Q,I,F,\delta)$, such that $\repspnrw{A} =
  \repspnrw{A_1} \cup \repspnrw{A_2}$. To this end, let $Q = Q_1 \cup Q_2$ be
  the set of states, $I,F: Q \to \srd$ with $I(q) = I_i(q)$ and $F(q) = F_i(q)$.
  Furthermore, let $\delta(p,a,q) = \delta_i(p,a,q)$ if $p,q \in Q_i$ 
  and $\delta(p,a,q) = \srzero$ if $p,q$ are not from the state set of the same
  automaton. We observe that this construction can be carried out in linear time.
  It remains to show the correctness of the construction. To this end, observe
  that every nonzero run $\rn$ of $A$ can only consist of states $q \in Q_1$ or $q
  \in Q_2$. Let $\doc \in \docs$ be an arbitrary document. The set
  $\Rn{A}{\doc}$ of all valid and nonzero runs of $A$ on $\doc$ is the union of
  two sets $P_1(A,\doc),P_2(A,\doc)$, where a run $\rn$ is in $P_i(A,\doc)$ if
  it consists of states in $Q_i$. Furthermore, it holds that $\rn \in
  P_i(A,\doc)$ if and only if $\rn \in \Rn{A_i}{\doc}$ and therefore,
  \begin{align*}
    \repspnrw{A}(\doc)(\tup)
    & =
    \boplus_{\rn\in \Rn{A}{\doc} \text{ and } \tup = \tup_{\rn}}\weight{\rn} \\
    & = 
    \left(\boplus_{\rn\in P_1(A,\doc) \text{ and } \tup = \tup_{\rn}}\weight{\rn}\right)
    \srplus
    \left(\boplus_{\rn\in P_2(A,\doc) \text{ and } \tup = \tup_{\rn}}\weight{\rn}\right) \\
    & =
    \left(\boplus_{\rn\in \Rn{A_1}{\doc} \text{ and } \tup = \tup_{\rn}}\weight{\rn}\right)
    \srplus
    \left(\boplus_{\rn\in \Rn{A_2}{\doc} \text{ and } \tup = \tup_{\rn}}\weight{\rn}\right) \\
    & = 
    \repspnrw{A_1}(\doc)(\tup)
    \srplus
    \repspnrw{A_2}(\doc)(\tup).
  \end{align*}
  This concludes the proof that $\repspnrw{A} = \repspnrw{A_1} \cup
  \repspnrw{A_2}$.
\end{proof}

\begin{lem}\label{lem:closureProjection}
	Given a $\K$-weighted \vset-automaton $A$ and a subset $X \subseteq V$ of the
  variables $V$ of $A$, there exists a weighted \vset-automaton $A^\prime$ with
  $\repspnrw{A^\prime} = \pi_X\repspnrw{A}$. Furthermore, if $A$ is functional,
  then $A^\prime$ can be constructed in polynomial time.
\end{lem}
\begin{proof}
  Let $A \df (\Sigma, V,Q,I,F,\delta)$ and $V^- = V \setminus X$. If $A$ is not
  yet functional, we can assume by Proposition~\ref{prop:functionalNormalform}
  that it is, at exponential cost in the number of variables of $A$.
  Furthermore, assume that, for every nonzero transition, there is a run $\rn$
  which uses the transition. Due to $A$ being functional, we will be able to
  construct $A^\prime$ by replacing all transitions labeled with a variable
  operation $o \in \varop{V^-}$ with an $\varepsilon$-transition of the same
  weight. More formally, let $A^\prime \df (\Sigma,X,Q,I,F,\delta^\prime)$, such
  that
  \begin{itemize}
  \item $\delta^\prime(p,o,q) = \delta(p,o,q)$ for all $p,q \in Q$ and $o \in
    \Sigma \cup \{\varepsilon\} \cup \varop{X}$, and
  \item $\delta^\prime(p,\varepsilon, q) = \delta(p,o,q)$ for all $p,q \in Q$
    and $o \in \varop{V^-}$.
  \end{itemize}

  We first argue why $\delta^\prime$ is well defined. Towards a contradiction,
  assume that $\delta^\prime$ is not well-defined. This can only happen if $A$
  has two transitions $\delta(p,o_1,q)$ and $\delta(p,o_2,q)$ with $o_1,o_2 \in
  \varop{V^-} \cup \{\varepsilon\}$ and $o_1 \neq o_2$. Therefore, there are two
  runs $\rn_1,\rn_2$ of A, which only differ on this transition, that is,
  $\rn_1$ uses $\delta(p,o_1,q)$ and $\rn_2$ uses $\delta(p,o_2,q)$
  respectively. Since $o_1 \neq
  o_2$ and $o_1,o_2 \in \varop{V^-} \cup \{\varepsilon\}$, either $\rn_1$ or
  $\rn_2$ are not valid, contradicting functionality of $A$.
  
  It remains to show that $\repspnrw{A^\prime} = \pi_X\repspnrw{A}$. To this
  end, let $\doc \in \docs$ be an arbitrary document. Every run $\rn$ of $A$
  selecting $\tup$ on $\doc$ corresponds to exactly one run $\rn'$ of $A'$
  selecting $\tup^\prime$ on $\doc$ such that $\tup^\prime = \tup \restriction
  X$ and $\weight{\rn} = \weight{\rn^\prime}$. Therefore,
  
  \begin{align*}
    \pi_X\repspnrw{A}(\doc)(\tup^\prime)
    & =   \boplus_{\tup^\prime = \tup \restriction X \text{ and } \repspnrw{A}(\doc)(\tup)\ne \srzero}
        \repspnrw{A}(\doc)(\tup) \\
    & =  
        \boplus_{\tup^\prime = \tup \restriction X \text{ and } \repspnrw{A}(\doc)(\tup)\ne \srzero}
        \quad
        \boplus_{\rn\in \Rn{A}{\doc} \text{ and } \tup = \tup_{\rn}} \weight{\rn} \\
    & = 
          \boplus_{\rn^\prime \in \Rn{A^\prime}{\doc} \text{ and } \tup^\prime = \tup_{\rn^\prime}}
        \weight{\rn^\prime} \\
    & = 
        \repspnrw{A^\prime}(\doc)(\tup^\prime).
  \end{align*}
  Therefore, $\repspnrw{A^\prime} = \pi_X\repspnrw{A}$.
\end{proof}

We will now show that regular annotators are closed under join. Freydenberger et
al.~\cite[Lemma 3.10]{DBLP:conf/pods/FreydenbergerKP18}, showed that, given two
functional $\B$-weighted \vset-automata $A_1$ and $A_2$, one can construct a
functional \vset-automaton $A$ with $\repspnrk{A}{\B} =\repspnrk{A_1}{\B} \join
\repspnrk{A_2}{\B}$ in polynomial time. The construction is based on the
classical product construction for the intersection of NFAs. However, $A_1$ and
$A_2$ can process consecutive variable operations in different orders which must
be considered during the construction. To deal with this issue, we adapt and
combine multiple constructions from the literature.

To be precise, we adopt so called \emph{extended \vset-automata} as
defined by Amarilli et al.~\cite{AmarilliBMN19} by adding weights to
the transitions.\footnote{We note that extended \vset-automata have
  been originally introduced in Florenzano et
  al.~\cite{FlorenzanoRUVV18}} An extended $\K$-weighted
\vset-automaton on alphabet $\Sigma$ and variable set $V$ is an
automaton $A_E = (\Sigma, V, Q, I, F, \delta)$, where
$Q = Q_v \uplus Q_\ell$ is a disjoint union of \emph{variable states
  $Q_v$} and \emph{letter states $Q_\ell$}.  Furthermore,
$I: Q \to \K$ is the initial weight function, such that
$I(q) = \srzero$, for every $q \in Q_\ell$. Analogously, $F: Q \to \K$
is a final weight function, such that $F(q) = \srzero$, for every
$q \in Q_v$. Finally, we define the (partial) transition function
$\delta: Q \times (\Sigma \cup 2^{\varop{V}}) \times Q \to \K$, such
that transitions labeled by $\sigma \in \Sigma$ originate in letter
states and terminate in variable states and $T \subseteq \varop{V}$
transitions are between variable states and letter states. More
formally, for every $\sigma \in \alphabet$, it holds that if
$\delta(p,\sigma,q) \neq \srzero$ then $p \in Q_\ell$ and $q \in
Q_v$. Furthermore, for every $T \subseteq \varop{V}$, if
$\delta(p,T,q) \neq \srzero$ then $p \in Q_v$ and $q \in Q_\ell$.

We define runs of extended weighted \vset-automata analogously to runs
on weighted \vset-automata. That is, a \e{run} $\rn$ of $A_E$ on a document
$\doc \df d_1 \cdots d_n$ is a sequence
\[
  (q_0, i_0) \overset{o_1}{\rightarrow} \cdots \overset{o_{m-1}}{\rightarrow}
  (q_{m-1}, i_{m-1}) \overset{o_{m}}{\rightarrow} (q_{m}, i_{m})
\]
where
\begin{itemize}
\item $I(q_0) \neq \srzero$ and $F(q_m) \neq \srzero$;
\item $i_0 = 1$, $i_m = n+1$, and $i_j \in \{1,\ldots, n\}$ for each $j \in
  \{1,\ldots,m-1\}$;
\item each $o_j$ is in $\Sigma \cup\{ \varepsilon \} \cup 2^{\varop{V}}$;
\item $i_{j+1} = i_j$ whenever $o_j \in \{\varepsilon \} \cup 2^{\varop{V}}$ and
  $i_{j+1} = i_j +1$, otherwise; 
\item $\delta(q_j,o_j,q_{j+1}) \ne \srzero$ for all $j\ge 0$.
\end{itemize}

The weight $\weight{\rn}$ of a run $\rn$ on an extended weighted
\vset-automaton, $\repspnrk{A_E}{\K}$, functionality, and unambiguity are
defined analogously to the weighted \vset-automata.

\begin{prop}\label{prop:toExtendedConversion}
  For every functional weighted \vset-automaton $A$, there exists an equivalent
  functional extended weighted \vset-automaton $A_E$ and vice versa. Given an
  automaton in one model, one can construct an automaton in the other model in
  polynomial time. Furthermore, the conversion preserves unambiguity.
\end{prop}
\begin{proof}
  Let $A \df (\alphabet, V,Q,I,F,\delta)$ be a weighted functional
  \vset-automaton.

  Due to Proposition~\ref{prop:functional} a weighted \vset-automaton is
  functional if and only if the automaton $A$ interpreted as $\B$-weighted
  \vset-automaton is functional. For functional \vset-automata it is well
  known\footnote{For example, compare
    Freydenberger~\cite{DBLP:journals/mst/Freydenberger19}, Freydenberger et
    al.~\cite{DBLP:conf/pods/FreydenbergerKP18}.} that there is a function $s:
  Q\times V \to\{w,o,c\}$, where
  \begin{itemize}
  \item $s(q,v) = w$ stands for ``waiting,'' meaning that no run $\rn$ of $A$ such that
    $\vop{v}$ is read before reaching state $q$.
  \item $s(q,v) = o$ stands for ``open,'' meaning that all runs $\rn$ of $A$ read
    $\vop{v}$ but not $\vcl{v}$ before reaching state $q$.
  \item $s(q,v) = c$ stands for ``closed,'' meaning that all runs $\rn$ of $A$ read
    $\vop{v}$ and $\vcl{v}$ before reaching state $q$.
  \end{itemize}
  Based on $s$, we define the function $S: Q \times Q \to
  \varop{V}$, such that $S(q,q') = T$, if on every run $\rn$ of $A$ which visits
  $q'$ after $q$, exactly the variable operations $T$ must be read between $q$
  and $q'$. More formally, $\vop{x} \in S(p,q)$ if and only if $s(p,x) = w$
  and $s(q,x) \neq w$ and $\vcl{x} \in S(p,q)$ if and only if $s(p,x) \neq c$
  and $s(q,x) = c$.

  We assume, \mbox{w.l.o.g.}, that the states of $A$ are $\{1,\ldots,n\}$
  for some $n \in \N$. For every state $i \in Q$, we define the vector $V_i$,
  where
  \[
    V_i(j) =
    \begin{cases}
      \srzero & \text{if }i\neq j\\
      \srone & \text{if } i=j\;.
    \end{cases}
  \]
  Furthermore, we define the $n\times n$ matrix $M_{p,q}$ where
  \[
    M_{p,q}(i,j) =
    \begin{cases}
      \delta(i,o,j) & \text{if } o \in S(p,q)\\
      \srone & \text{otherwise}\;.
    \end{cases}
  \]
  
  We construct the weighted extended functional \vset-automaton $A_E \df
  (\alphabet, V, Q_\ell \cup Q_v, I_E, F_E, \delta_E)$ as follows. Let $Q_\ell \df
  \{q_\ell \mid q \in Q\}$ and $Q_v \df \{q_v \mid q \in Q\}$ be two disjoint
  copies of the states of $A$. Furthermore, let
  \begin{align*}
    I_E(q) &\df
    \begin{cases}
      I(q) & \;\text{if } q \in Q_v\\
      \srzero & \;\text{if } q \in Q_\ell\;;
    \end{cases}\\
    F_E(q) &\df
    \begin{cases}
      \srzero & \text{if } q \in Q_v\\
      F(q) & \text{if } q \in Q_\ell\;.
    \end{cases}
  \end{align*}
  We define $\delta_E$ as follows
  \[
    \begin{array}{r@{\ }ll}
      \delta_E(p_l,\sigma,q_v)
      & \df \delta(p,\sigma,q)
      & \text{for all } \sigma \in \alphabet,\\
      \delta_E(p_v,O,q_\ell)
      & \df V_{p_v} \times (M_{p_v,q_\ell})^{|O|} \times V_{q_\ell}^T
      & \text{for } O = S(p_v,q_\ell)\\
      \delta_E(p_v,\emptyset,p_\ell)
      &\df \srone
    \end{array}
  \]

  We observe that per assumption that $\srd$ has an efficient encoding, it
  follows that $A_E$ can be constructed in polynomial time. It remains to show
  that $\repspnrk{A}{\K} = \repspnrk{A_E}{\K}$. To this end, we define a
  function, which maps valid runs of $A$ to runs of $A_E$. More formally, let
  \[
    (q_0, i_0) \overset{o_1}{\rightarrow} \cdots \overset{o_{m-2}}{\rightarrow} (q_{m-1}, i_{m-1})
    \overset{o_{m-1}}{\rightarrow}  (q_{m}, i_{m}) 
  \]
  be a run $\rn$ of $A$ on $\doc = \doc_1 \cdots \doc_n$.

  Let $q_v^1 \in Q_v$ (resp., $q_\ell^{n+1} \in Q_\ell$) be the variable state
  (resp., letter state) corresponding to $q_0$ (resp., $q_{m}$). Furthermore,
  for $1 \leq k \leq n$, let $q_\ell^{k},q_v^{k+1}$ be the states corresponding
  to the states visited by $\rn$ while reading the symbol $\doc_{k}$. That is,
  for $(q_j,k) \overset{\doc_{k}}{\rightarrow} (q_{j+1},k+1)$ in $\rn$, $q_\ell^{k-1}$
  corresponds to $q_j$ and $q_v^{k}$ to $q_{j+1}$.
  
  We define $f(\rn)$ as the run $\rn_E \in \Rn{A_E}{\doc}$ such that
  \[
    \rn_E = (q_v^1,1) \xrightarrow{S(q_v^1,q_\ell^1)} (q_\ell^1,1)
    \xrightarrow{\doc_1} (q_v^2,2) \xrightarrow{S(q_v^2,q_\ell^2)} \cdots
    \xrightarrow{\doc_n} (q_v^{n+1},n+1) \xrightarrow{S(q_v^{n+1},q_\ell^{n+1})}
    (q_\ell^{n+1},n+1)\;.
  \]

  For every valid run $\rn_E \in \Rn{A_E}{\doc}$, it holds that
  $\weight{\rn_E} = \boplus_{\rn \in \Rn{A}{\doc} \text{ with } f(\rn) = \rn_E}
  \weight{\rn}$. Therefore, it follows that
  \begin{align*}
    \repspnrw{A_E}(\doc)(\tup)
    = & \boplus_{\rn_E\in \Rn{A_E}{\doc} \text{ and } \tup = \tup_{\rn_E}} \weight{\rn_E} \\
    = & \boplus_{\rn_E\in \Rn{A_E}{\doc} \text{ and } \tup = \tup_{\rn_E}} \
        \boplus_{\rn\in \Rn{A}{\doc} \text{ with } f(\rn) = \rn_E} \weight{\rn}\\
    = & \boplus_{\rn\in \Rn{A}{\doc} \text{ and } \tup = \tup_{\rn}} \weight{\rn}\\
    = & \;\; \repspnrw{A}(\doc)(\tup)\;.
  \end{align*}
  It remains to show that $A_E$ is unambiguous if $A$ is unambiguous. To this
  end, assume that $A_E$ is not unambiguous. Thus, there must be two runs
  $\rn_E^1 \neq \rn_E^2$ on $A_E$, encoding the same tuple. By construction of
  $A_E$, there must be two runs $\rn_1 \neq \rn_2$ of $A$ which encode the same
  tuple, however this contradicts the unambiguity of $A$. Therefore $A_E$ must
  be unambiguous.
    
  For the other direction, one can construct a weighted \vset-automaton $A$ with
  $\varepsilon$-transitions,\footnote{By Proposition~\ref{prop:epsilonRemoval}
    the $\varepsilon$-transitions can be removed in polynomial time.} by
  replacing every edge $\delta(p,T,q) = w$ by a sequence of transitions
  $\delta(p,v_1,q_1) = w$, $\delta(q_1,v_2,q_2) = \srone, \ldots,
  \delta(q_{n-1},v_n,q) = \srone$, where $T = \{v_1,\ldots,v_n\}$ and
  $q_1,\ldots,q_{n-1}$ are new states. We observe that only the first transition
  has weight $w$, whereas all other transitions have weight $\srone$. This
  construction also runs in polynomial time and it is straightforward to verify
  that $\repspnrk{A}{\K} = \repspnrk{A_E}{\K}$ and that $A$ is unambiguous if
  $A_E$ is unambiguous.
\end{proof}

\begin{prop}\label{prop:extendedJoin}
  Let $A_1,A_2$ be two functional extended $\K$-weighted \vset-automata. One can
  construct a functional extended $\K$-weighted \vset-automaton $A$ in
  polynomial time, such that $\repspnrw{A} = \repspnrw{A_1} \join
  \repspnrw{A_2}$. Furthermore, $A$ is unambiguous if $A_1$ and $A_2$ are
  unambiguous.
\end{prop}
\begin{proof}
  Let $A_1 = (\alphabet, V_1, Q_1, I_1, F_1, \delta_1)$ and $A_2 = (\alphabet,
  V_2, Q_2, I_2, F_2, \delta_2)$ be two $\K$-weighted extended functional
  \vset-automata. We construct a $\K$-weighted extended functional
  \vset-automaton $A = (\alphabet, V_1 \cup V_2, Q_1 \times Q_2, I, F, \delta)$
  such that $\repspnrw{A} = \repspnrw{A_1} \join \repspnrw{A_2}$. To this end,
  let $I(q_1,q_2) = I_1(q_1) \srtimes I_2(q_2)$ and $F(q_1,q_2) = F_1(q_1)
  \srtimes F_2(q_2)$. Furthermore, let
  \[
    \delta((p_1,p_2), \sigma, (q_1,q_2)) = \delta_1(p_1,\sigma,q_1) \srtimes
    \delta_2(p_2,\sigma,q_2)\;,
  \]
  if $\sigma \in \alphabet$, and otherwise, if $T \subseteq \varop{V}$,
  \[
    \delta\big((p_1,p_2), T, (q_1,q_2)\big) = \delta_1\big(p_1,T \cap
    \varop{V_1},q_1\big) \srtimes \delta_2(p_2,T\cap\varop{V_2},q_2)\;.
  \]
  We observe that $A$ can be constructed in polynomial time. We have to show
  that $\repspnrw{A} = \repspnrw{A_1} \join \repspnrw{A_2}$. Let $\doc \in
  \docs$ be a document and $\tup$ be a tuple. Every run $\rn \in \Rn{A}{\doc}$
  with $\tup_\rn = \tup$ originates from of a set of runs $\rn_1 \in
  \Rn{A_1}{\doc}$ selecting $\pi_{V_1}\tup$ and a set of runs $\rn_2 \in
  \Rn{A_2}{\doc}$ selecting $\pi_{V_2}\tup$. Due to distributivity of $\srtimes$
  over $\srplus$, it holds that $\weight{\rn} = \weight{\rn_1} \srtimes
  \weight{\rn_2}$. Furthermore, every run in $A$ corresponds to exactly one run
  in $A_1$ and one run in $A_2$. It follows directly that
  $\repspnrw{A}(\doc)(\tup) = \repspnrw{A_1}(\doc)(\pi_{V_1}\tup) \srtimes
  \repspnrw{A_2}(\doc)(\pi_{V_2}\tup)$ and that the construction preserves
  unambiguity.
\end{proof}
  
We now show that regular annotators are closed under join.
\begin{lem}\label{lem:closureJoin}
  Given two $\K$-weighted \vset-automata $A^1$ and $A^2$, one can
  construct a weighted functional \vset-automaton $A$ with
  $\repspnrw{A} = \repspnrw{A^1} \join \repspnrw{A^2}$. Furthermore,
  $A$ can be constructed in polynomial time if $A^1$ and $A^2$ are
  functional and $A$ is unambiguous if $A_1$ and $A_2$ are
  unambiguous.
\end{lem}
\begin{proof}
  If $A^1$ and $A^2$ are not yet functional, we can assume that they are at an
  exponential cost in their number of variables (cf.
  Proposition~\ref{prop:functionalNormalform}). By
  Proposition~\ref{prop:toExtendedConversion}, one can construct functional
  extended weighted \vset-automata $A_E^1, A_E^2$ with $\repspnrk{A^i}{\K} =
  \repspnrk{A_E^i}{\K}$. Furthermore, due to
  Proposition~\ref{prop:extendedJoin}, one can construct a functional extended
  weighted \vset-automaton $A_E$ with $\repspnrk{A_E}{\K} = \repspnrk{A_E^1}{\K}
  \join \repspnrk{A_E^2}{\K}$. Thus, again applying
  Proposition~\ref{prop:toExtendedConversion}, one can construct a functional
  weighted \vset-automaton $A_E$ with $\repspnrk{A_E}{\K} = \repspnrk{A_E^1}{\K}
  \join \repspnrk{A_E^2}{\K} = \repspnrk{A^1}{\K} \join \repspnrk{A^2}{\K}$.
  Note that all constructions are in polynomial time if $A^1$ and $A^2$ are
  functional and preserve unambiguity. Thus, concluding the proof with $A \df
  A_E$.
\end{proof}

The previous lemma also has applications to unambiguous functional
\vset-automata over the Boolean semiring.

\begin{cor}\label{cor:ufvsaJoin}
  Given two unambiguous functional \vset-automata $A_1,A_2$ over the Boolean
  semiring, one can construct an unambiguous functional \vset-automaton $A$ with
  $\repspnrk{A}{\B} = \repspnrk{A_1}{\B} \join \repspnrk{A_2}{\B}$ in polynomial
  time.
\end{cor}

 \subsection{Closure Under String Selection}
A $k$-ary string relation is \emph{recognizable} if it is a finite union of
Cartesian products $L_1 \times \cdots \times L_k$, where each $L_i$ is a regular
language over $\Sigma$~\cite{sakarovitch_2009}. Let $\reg{\K}$ be the set of
regular $\K$-annotators. We say that a $k$-ary string relation\footnote{Recall
  that a \emph{$k$-ary string relation} is a subset of $\docs^k$.} $R$ is
\emph{selectable by regular $\K$-annotators} if the class of $\srd$-annotators
is closed under the string selection $\sigma^R$. More formally:
\[
  \{\sigma^R_{x_1,\ldots,x_k}(S) \mid S \in \regk \text{ and } x_i \in
  \vars(S) \text{ for all } 1 \leq i \leq k\} \subseteq \regk\ ,
\]
that is, the class of $\K$-annotators is closed under selection using $R$. If
$\K = \B$, we say that $R$ is \emph{selectable by document spanners}. Fagin et
al.~\cite{DBLP:journals/jacm/FaginKRV15} proved that a string relation is
recognizable \emph{if and only if} it is selectable by document spanners. Here,
we generalize this result in the context of weights and annotation. Indeed, it
turns out that the equivalence is maintained for all positive semirings.

\begin{thm}\label{thm:positiveSelect}
  Let $\sr$ be a positive semiring and $R$ be a string relation. The following
  are equivalent:
  \begin{enumerate}
  \item $R$ is recognizable.
  \item $R$ is selectable by document spanners.
  \item $R$ is selectable by $\K$-annotators.
  \end{enumerate}
\end{thm}
\begin{proof}
  The equivalence between (1) and (2) is known~\cite[Theorem
  4.16]{DBLP:journals/jacm/FaginKRV15}. We will show the implication (2)
  $\Rightarrow$ (3) in Lemma~\ref{lem:select} and implication (3) to (2) in
  Lemma~\ref{lem:positiveSelect}.
\end{proof}

As we will see, the proof of Lemma~\ref{lem:select} is heavily based on the
closure properties from Theorem~\ref{theo:closed-algebra} and holds beyond
positive semirings. For the proof of Lemma~\ref{lem:positiveSelect}, we use
semiring morphisms to turn $\K$-weighted \vset-automata into $\B$-weighted
\vset-automata and need positivity of the semiring. We need some preliminary
results in order to give the proofs of the lemma.

\begin{defi}\label{def:stringrelationselector}
  Let $R$ be a $k$-ary string relation. A $\K$-weighted \vset-automaton $A_R^\K$
  with variables $\{x_1, \ldots, x_k\}$ selects $R$ over $\K$ if for every
  document $\doc \in \docs$ and every tuple $\tup$ it holds that
  $\repspnrk{A_R^\K}{\K}(\doc)(\tup) = \srone$ if
  $(\doc_{\tup(x_1)},\ldots,\doc_{\tup(x_k)}) \in R$, and $\srzero$, otherwise,
  if $(\doc_{\tup(x_1)},\ldots,\doc_{\tup(x_k)}) \notin R$.
\end{defi}

\begin{lem}\label{lem:selectableRelationSpanner}
  Let $R$ be a $k$-ary string relation. Then $R$ is selectable by
  $\K$-annotators if and only if there is a \vset-automaton $A_R^\K$ that
  selects $R$ over $\K$.
\end{lem}
\begin{proof}
  Assume that $R$ is selectable by $\K$-annotators. Let $A$ be the $\K$-weighted
  \vset-automaton that assigns weight $\srone$ to all possible tuples for all
  documents. As $R$ is selectable by $\K$-annotators,
  $\sigma^R_{x_1,\ldots,x_k}(\repspnrk{A}{\K})$ must be a regular
  $\K$-annotator. Thus, the $\K$-weighted \vset-automaton $A_R^\K$ representing
  $\sigma^R_{x_1,\ldots,x_k}(\repspnrk{A}{\K})$ selects $R$ over $\K$.

  For the other direction, let $A_R^\K$ be as defined. Let $A$ be a $\K$-weighted
  \vset-automaton. Per definition of string selection,
  $\sigma^R_{x_1,\ldots,x_k}(\repspnrk{A}{\K})(\tup) = \repspnrk{A}{\K}(\tup)
  \srtimes \srzero = \srzero$ if $(\doc_{\tup(x_1)},\ldots,\doc_{\tup(x_k)})
  \notin R$ and $\sigma^R_{x_1,\ldots,x_k}(\repspnrk{A}{\K})(\tup) =
  \repspnrk{A}{\K}(\tup) \srtimes \srone = \repspnrk{A}{\K}(\tup)$, otherwise.
  Therefore, $\sigma_{x_1,\ldots,x_k}^R(\repspnrk{A}{\K}) = \repspnrk{A}{\K}
  \join \repspnrk{A_R^\K}{\K}$, which proves that $R$ is selectable by
  $\K$-annotators, as $\K$-annotators are closed under join (c.f.
  Theorem~\ref{theo:closed-algebra}).
\end{proof}

We will now define means of transferring the structure of weighted automata
between different semirings, that is, we define \emph{$\B$-projections} and
\emph{$\K$-extensions} of weighted \vset-automata.

\begin{defi}\label{def:bprojection}
  Let $A$ be a weighted \vset-automaton over $\K$. A $\B$-weighted
  \vset-automaton $A^\B$ is a \emph{$\B$-projection of $A$} if, for every
  document $\doc \in \docs$, it holds that $\tup \in \repspnrk{A^\B}{\B}(\doc)
  \Leftrightarrow \tup \in \repspnrk{A}{\srd}(\doc)$.
\end{defi}

\begin{defi}\label{def:kextension}
  Let $A$ be a $\B$-weighted \vset-automaton. Then a $\K$-weighted
  \vset-automaton $A^\K$ is called a \emph{$\K$-extension of $A$} if, for every
  document $\doc \in \docs$ and every tuple $\tup$, the following are
  equivalent:
  \begin{enumerate}
  \item $\tup \in \repspnrk{A}{\B}(\doc)$
  \item $\tup \in \repspnrk{A^\K}{\K}(\doc)$ and
    $\repspnrk{A^\K}{\K}(\doc)(\tup) = \srone$.
  \end{enumerate}
  Furthermore, $A^\K$ has exactly one run for every tuple in
  $\repspnrk{A^\K}{\srd}(\doc)$.
\end{defi}

We now show that a $\B$-projections of a $\K$-weighted \vset-automaton $A$
exists if $\K$ is positive. Furthermore, a $\K$-extensions of a $\B$-weighted
$\vset$-automaton always exists. To this end, let $\sr$ and $\srp$ be semirings.
For a function $f: \srd \to \srdp$ and a weighted \vset-automaton $A \df (\Sigma,
V,Q,I,F,\delta)$ over $\srd$, we define the weighted \vset-automaton $A_f \df
(\Sigma, V,Q,I_f,F_f,\delta_f)$ over $\srd^\prime$, where $I_f \df f \circ I,
F_f \df f \circ F$, and $\delta_f \df f \circ \delta$.

\begin{lem}\label{lem:SrdToB}
  Let $\K$ be a positive semiring. Then there exists a $\B$-projection $A^\B$ of
  $A$ for every $\K$-weighted \vset-automaton $A$.
\end{lem}
\begin{proof}
  Let $\morphtob: \srd \to \B$ be the function
  \[
    \morphtob(x) =
    \begin{cases}
      \true & \text{if } x \neq \srzero,\\
      \false & \text{if } x = \srzero.
    \end{cases}
  \]
  Eilenberg~\cite[Chapter VI.2]{Eilenberg74} showed that, due to $\K$ being
  positive\footnote{Eilenberg~\cite[Chapter VI.2]{Eilenberg74} actually showed
    that $\morphtob$ is a semiring morphism if and only if $\K$ is positive.},
  the function $\morphtob$ is a semiring morphism, that is,
  \[
    \begin{array}{rl@{\ \ }rl}
      \morphtob(x \srplus^\K y) = \morphtob(x) \srplus^\B \morphtob(y), & \morphtob(\srzero) = \false, \\
      \morphtob(x \srtimes^\K y) = \morphtob(x) \srtimes^\B \morphtob(y), & \morphtob(\srone) = \true.
    \end{array}
  \]
  Observe that these properties ensure that,
  for every document $\doc \in \docs$ and every tuple $\tup \in
  \repspnrk{A}{\K}$, it holds that
  \[
    \morphtob\left(\repspnrk{A}{\K}(\doc)(\tup)\right) = \repspnrk{A_\morphtob}{\Kp}(\doc)(\tup). 
  \]
  Therefore, $A_\morphtob$ is a $\B$-projection of $A$.
\end{proof}

\begin{lem}\label{lem:BToSrd}
  Every $\B$-weighted \vset-automaton $A$ has a $\K$-extension. 
\end{lem}
\begin{proof}
  Let $A \df (V,Q,I,F,\delta)$ be a $\B$-weighted \vset-automaton. Doleschal et
  al.~\cite[Proposition 4.4]{DoleschalKMNN19} showed that, for every
  \vset-automaton $A$, there is an equivalent deterministic \vset-automaton
  $A_{\mathtt{det}}$. Note that a deterministic \vset-automaton has exactly one
  run for every tuple $\tup \in \repspnrk{A_{\mathtt{det}}}{\B}(\doc)$.
  Therefore, \mbox{w.l.o.g.}, we can assume that $A$ has this property. Let
  $\funcfromb: \B \to \K$ be the function\footnote{Notice that $\funcfromb$ is
    not necessarily a semiring morphism. Depending on $\K$, it may be the case
    that $\srone \srplus \srone = \srzero$, contradicting the properties of
    semiring morphisms. Take $\K = \Z / 2\Z$, for instance.}
  \[
    \funcfromb(x) =
    \begin{cases}
      \srone  & \text{if }x = \true, \\
      \srzero & \text{if }x = \false.
    \end{cases}
  \]
  Observe that $A_\funcfromb$ also has exactly one run for every tuple $\tup \in
  \repspnrk{A_{\mathtt{det}}}{\K}(\doc)$. It remains to show that $A_\funcfromb$
  is indeed a $\K$-extension of $A$. To this end, let $\doc \in \docs$ be a
  document. We now show the equivalence between (1) and (2).

  (1) implies (2): Let $\tup \in \repspnrk{A}{\B}(\doc)$. Per assumption, $A$ has
  exactly one run $\rn$ on $\doc$ for $\tup$. Let $\rn_\funcfromb \df
  \funcfromb(\rn)$ be the run, resulting from $\rn$ by replacing all weights $w$
  with $\funcfromb(w)$. Observe that $\rn_\funcfromb$ must be a run of
  $A_\funcfromb$ on $\doc$ accepting $\tup$. Per construction, all transitions
  of $A_\funcfromb$ have weight $\srzero$ or $\srone$. Thus, (2) must hold.

  (2) implies (1): Let $\tup \in \repspnrk{A_\funcfromb}{\K}(\doc)$ and
  $\repspnrk{A^\prime}{\K}(\doc)(\tup) = \srone$. Thus, there is a run
  $\rn_\funcfromb$ of $A_\funcfromb$ on $\doc$ accepting $\tup$. Therefore,
  there must also be a run $\rn$ of $A$ on $\doc$, accepting $\tup$,
  concluding the proof.
\end{proof}

We are now ready to prove the two main results of this section.
\begin{lem}\label{lem:select}
  Let $R$ be a string relation, which is selectable by document spanners.
  Then $R$ is also selectable by $\srd$-Annotators.
\end{lem}
\begin{proof}
  Let $A$ be a $\K$-weighted \vset-automaton and $R$ be a relation that is
  selectable by regular $\B$-annotators. We have to show that every string
  selection $\sigma^R_{x_1,\dots,x_k}\repspnrk{A}{\K}$ is definable by a
  $\K$-weighted \vset-automaton. By assumption $R$ is selectable by regular
  $\B$-annotators. Let $A_R^\B$ be the \vset-automaton that selects $R$ over
  $\B$, which exists by Lemma~\ref{lem:selectableRelationSpanner}. Let $A_R^\K$
  be a $\K$-extension of $A_R^\B$ \vset-automaton, which exists by
  Lemma~\ref{lem:BToSrd}. Thus, $A_R^\K$ selects $R$ over $\K$ and therefore, by
  Lemma~\ref{lem:selectableRelationSpanner}, $R$ is selectable by
  $\srd$-Annotators.
\end{proof}

\begin{lem}\label{lem:positiveSelect}
  Let $\sr$ be a positive semiring and $R$ be a string relation, which is
  selectable by $\srd$-Annotators. Then $R$ is also selectable by document
  spanners.
\end{lem}

\begin{proof}
  Let $R$ be a string relation selectable by $\K$-annotators and let $A$ be a
  $\B$-weighted \vset-automaton. We have to show that $R$ is also selectable
  over $\B$, i.e., there is a $\B$-weighted \vset-automaton $A_R^\B$ such that
  $\repspnrk{A_R^\B}{\B} = \sigma^R_{x_1,\dots,x_k}\repspnrk{A}{\B}$. Let
  $A^{\K}$ be a $\K$-extension of $A$, which exists by Lemma~\ref{lem:BToSrd}.
  Per assumption $R$ is selectable over $\srd$, therefore, due to
  Lemma~\ref{lem:selectableRelationSpanner}, there exists a $\K$-weighted
  \vset-automaton $A_R^\K$ which selects $R$ over $\K$. Thus,
  $\sigma^R_{x_1,\dots,x_k}\repspnrk{A^{\K}}{\srd} = \repspnrk{A^\K_{R}}{\srd}$.
  Let $A^\B_R$ be a $\B$-projection of $A^\K_R$, which exists by
  Lemma~\ref{lem:SrdToB}. It remains to show that
  $\sigma^R_{x_1,\dots,x_k}\repspnrk{A}{\B} = \repspnrk{A^\B_R}{\B}$. Let $\tup
  \in \repspnrk{A^\B_R}{\B}$. By Lemma~\ref{lem:SrdToB}, $\tup \in
  \repspnrk{A^\K_R}{\srd}$ and therefore, $\tup \in
  \sigma^R_{x_1,\dots,x_k}\repspnrk{A^{\srd}}{\srd}$. Per definition of string
  selection, it follows that $(\doc_{\tup(x_1)},\ldots,\doc_{\tup(x_k)}) \in R$
  and $\tup \in \repspnrk{A_{\srd}}{\srd}$. By Lemma~\ref{lem:BToSrd}, it
  follows that $(\doc_{\tup(x_1)},\ldots,\doc_{\tup(x_k)}) \in R$ and $\tup \in
  \repspnrk{A}{\B}$, and therefore $\tup \in
  \sigma^R_{x_1,\dots,x_k}\repspnrk{A}{\B}$. Observe that all implications in
  the previous argument where actually equivalences. Therefore, the inclusion
  $\sigma^R\repspnrk{A}{\B} \subseteq \repspnrk{A^\B_R}{\B}$ also holds.
\end{proof}

Observe that for the proof of Theorem~\ref{thm:positiveSelect} we only required
positivity of the semiring for the implication from (3) to (2). This raises the
question whether the equivalence can be generalized even further. We show next
that this is indeed the case, such as for the {\L}ukasiewicz semiring, which is
not positive.

\subsubsection{Beyond Positive Semirings}\label{sec:stringsele:beyondpositive}

We provide some insights about the cases where $\K$ is not positive. Recall
that, by Lemma~\ref{lem:select}, every string relation $R$, which is selectable
by document spanners is also selectable by $\K$-Annotators. Therefore, the
question is: for which semirings $\K$ does selectability by $\K$-annotators
imply selectability by ordinary document spanners? It turns out that this is
indeed possible for some non-positive semirings, such as the {\L}ukasiewicz
semiring \L.

Let $\srp$ be a subsemiring of a semiring $\K$.\footnote{Recall that a
  subsemiring of $\srd$ is a set $\srdp$, closed under addition and
  multiplication.} This semiring is \emph{minimal} if there is no
subsemiring of $\sr$ with fewer elements. Recall that a semiring $\K$ is
bipotent, if $a \srplus b \in \{a,b\}$, for every $a,b \in \K$. We begin with
some intermediate results.

\begin{lem}\label{lem:bipotentPositive}
  Let $\sr$ be a bipotent semiring. Then $\srd_{\min} \df \{\srzero, \srone\}$
  is the unique minimal subsemiring of $\srd$. Furthermore, $\srd_{\min}$ is
  isomorphic to the Boolen semiring.
\end{lem}
\begin{proof}
  For every semiring it holds that $\srzero \srtimes \srone = \srzero$, $\srone
  \srtimes \srzero = \srzero$, and $\srzero \srtimes \srzero = \srzero$.
  Furthermore, $\srone \srplus \srzero = \srone$ and $\srzero \srplus \srzero =
  \srzero$. As $\K$ is bipotent, it also holds that $\srone \srplus \srone =
  \srone$. Recall that a subsemiring of $\srd$ is a set $\srdp$, closed under
  addition and multiplication. Let $\srdp = \{\srzero, \srone\}$. Thus,
  $\K_{\min} = \{\srzero, \srone\}$ is a subsemiring of $\srd$, as
  $\{\srzero,\srone\}$ is closed under addition and multiplication. Observe that
  $\K_{\min}$ must be unique and minimal, as every subsemiring must at least
  contain $\srzero$ and $\srone$.
  
  It remains to show that $\K_{\min}$ is isomorphic to $\B$. To this end, let
  $\morphtob: \K_{\min} \to \B$ be the bijection 
  \[
    \morphtob(x) =
    \begin{cases}
      \true & \text{if } x = \srone,\\
      \false & \text{if } x = \srzero.
    \end{cases}
  \]
  It is straightforward to verify that $\morphtob$ is indeed a semiring
  isomorphism.
\end{proof}
It follows directly that $\K_{\min}$ is a positive semiring.

\begin{lem}\label{thm:bipotentSelect}
  Let $\K$ be a bipotent semiring such that $a \srtimes b = \srone$ implies that
  $a = b = \srone$. Then a string relation $R$ is selectable by $\K$-annotators
  if and only if it is selectable by $\K_{\min}$-annotators.
\end{lem}
\begin{proof}
  Every $\K_{\min}$-annotator is also a $\K$-annotator. Therefore, we only have to
  show that every string relation selectable by $\K$-annotators is also selectable
  by $\K_{\min}$-annotators.
  
  Let $A$ be a $\K_{\min}$-weighted \vset-automaton and $R$ be selectable by
  $\K$-annotators. We have to show that
  $\sigma^R_{x_1,\dots,x_k}\repspnrk{A}{\K_{\min}}$ is definable by a
  $\K_{\min}$-weighted \vset-automaton. Let $\doc \in \docs$ be a document.

  Per assumption $R$ is selectable over $\K$. Let $A_{\K}^R$ be a $\K$-weighted
  \vset-automaton, guaranteed by Lemma~\ref{lem:selectableRelationSpanner}.
  Therefore, $A_\K^R$ only assigns weight $\srone$ and $\srzero$. Recall that
  $\K$ is bipotent, that is, for every $a,b \in \K$, it holds that $a \srplus b
  \in \{a,b\}$. Therefore, for every $\tup \in \repspnrk{A_\K^R}{\K}(\doc)$
  there must be a run $\rn $ of $A_\K^R$ on $\doc$ with $\weight{\rn} = \srone$.
  Furthermore, as $a \srtimes b = \srone$ implies that $a = b = \srone$, this
  run must not have an edge with weight $a \neq \srone$. The existence a run
  $\rn$ of $A_\K^R$ on $\doc$ with weight $\weight{\rn} = \srone$ also implies
  that $\tup_\rn \in \repspnrk{A_\K^R}{\K}(\doc)$. Thus, there is a run $\rn$ of
  $A_\K^R$ on $\doc$ consisting only of edges with weight $\srone$ if and only
  if $\tup_\rn \in \repspnrk{A_\K^R}{\K}$. Therefore, we can assume,
  \mbox{w.l.o.g.}, that all edges in $A_\K^R$ have either weight $\srzero$ or
  $\srone$. Thus, $A_\K^R$ is a $\K_{\min}$-annotator and it follows from
  Lemma~\ref{lem:selectableRelationSpanner} that $R$ is selectable over
  $\K_{\min}$.
\end{proof}
 
\begin{cor}\label{cor:nonposSemiringRationalRelations}
  Let $\K$ be a bipotent semiring, such that $a \srtimes b = \srone$ implies
  that $a = b = \srone$. A string relation $R$ is recognizable if and only if it is
  selectable by $\K$-annotators.
\end{cor}
\begin{proof}
  Follows directly from
  Theorems~\ref{thm:positiveSelect},~\ref{thm:bipotentSelect}, and
  Lemma~\ref{lem:bipotentPositive}.
\end{proof}

Recall the \L ukasiewcz semiring, whose domain is $[0,1]$, with addition given
by $x \srplus y = \max(x,y)$, with multiplication $x \srtimes y = \max(0, x + y
- 1)$, zero element $0$, and one element $1$. Thus, for every $a,b \in [0,1]$,
$a \srplus b \in \{a,b\}$ and $a \srtimes b = 1$ if and only if $a = b = 1$.
Therefore, the {\L}ukasiewcz semiring satisfies the conditions of
Corollary~\ref{cor:nonposSemiringRationalRelations}.

\begin{cor}
  Let \L\xspace be the \L ukasiewcz semiring. A string relation $R$ is recognizable
  if and only if it is selectable by \L-annotators.
\end{cor}
 
\section{Evaluation Problems}\label{sec:eval}
We consider two types of evaluation problems in this section: \emph{answer
  testing} and \emph{best weight evaluation}. The former is given an annotator,
document $\doc$, and tuple $\tup$, and computes the annotation of $\tup$ in
$\doc$ according to the annotator. The latter does not receive the tuple as
input, but receives a weight threshold and is asked whether there exists a tuple
to which a weight greater than or equal to the threshold is assigned.

\subsection{Answer Testing}

It follows from Freydenberger~\cite[Lemma 3.1]{DBLP:journals/mst/Freydenberger19}
that answer testing is \np-complete for $\B$-weighted \vset-automata in general.
Indeed, he showed that, given a $\B$-weighted \vset-automaton $A$, it is
\np-complete to check if $A$ returns an output on the empty document, so it is
even \np-complete to check if the tuple of empty spans is returned or not.
However, the proof makes extensive use of non-functionality of the automaton.
Indeed, we can prove that answer testing is tractable for functional weighted
\vset-automata.
\begin{thm}\label{thm:weightedTupleEvalPtime}
  Given a functional weighted \vset-automaton $A$, a document $\doc$, and a tuple
  $\tup$, the weight $\repspnrw{A}(\doc)(\tup)$ assigned to
  $\tup$ by $A$ on  $\doc$ can be computed in polynomial time.
\end{thm}
\begin{proof}
  Let $A$, $\doc$, and $\tup$ be as stated. Per definition, the weight assigned
  to $\tup$ by $A$ is 
  \[
    \repspnrw{A}(\doc)(\tup) \df \boplus_{\rn\in \Rn{A}{\doc} \text{ and } \tup
      = \tup_{\rn} } \weight{\rn}.
  \]
  Therefore, in order to compute the weight $\repspnrw{A}(\doc)(\tup)$, we need
  to consider the weights of all runs $\rn$ for which $\tup = \tup_{\rn}$.
  Furthermore, multiple runs can select the same tuple $\tup$ but assign
  variables in a different order.\footnote{This may happen when variable
    operations occur consecutively, that is, without reading an alphabet symbol
    in between.}
    
  Let $A_E$ be the functional extended weighted \vset-automaton
  corresponding to $A$, as constructed by
  Proposition~\ref{prop:toExtendedConversion}. It follows that
  $\repspnrw{A}(\doc)(\tup) = \repspnrw{A_E}(\doc)(\tup)$. Let $\rn$
  be a run of $A_E$ on $\doc$ and $\tup$, and let $\doc_\tup$ be the
  document obtained from $\rn$ by concatenating the labels of the
  transitions of $\rn$.\footnote{We note that we assume in the
    following that every set $X \subseteq \varop{V}$ is represented by
    an unique label that is not in $\Sigma$. That is, we can assume
    w.l.o.g.\ that no set of the form $X \subseteq \varop{V}$ is an
    element from $\Sigma$ and just use subsets of $\varop{V}$ as
    labels.}  Observe that $\doc_\tup$ encodes $\doc$ and $\tup$ and
  is uniquely defined by $\doc$ and $\tup$.
  
  It follows that, if $A_E$ is interpreted as an weighted automaton,
  the weight $\repspnrw{A_E}(\doc)(\tup)$ is exactly the weight which
  is assigned by $A_E$ to the input word $\doc_\tup$.
  
  We note that computing the weight assigned by a weighted automaton
  on an input word $w$ strongly depends on the cost model and the used
  semiring. We therefore give an explicit proof that the weight can be
  computed in polynomial time if the semiring has an efficient
  encoding.

  We define an functional extended weighted \vset-automaton $A_\tup$,
  such that $\repspnrw{A_\tup}(\doc)(\tup) = \srone$ and
  $\repspnrw{A_\tup}(\doc)(\tup^\prime) = \srzero$ for all
  $\tup^\prime \neq \tup$. Such an automaton $A_\tup$ can be defined
  using a chain of $|\doc_\tup| + 1$ states, which checks that the
  input document is $\doc$ and which has exactly one nonzero run
  $\rn$, with $\weight{\rn} = \srone$ and $\tup_\rn = \tup$.
  
  By Proposition~\ref{prop:extendedJoin}, there is a weighted \vset-automaton
  $A^\prime$ such that $\repspnrw{A^\prime} = \repspnrw{A_E} \join
  \repspnrw{A_\tup}$. It follows directly from the definition of $A^\prime$ that
  $\repspnrw{A^\prime}(\doc^\prime)(\tup^\prime) = \srzero$ if $\doc^\prime \neq
  \doc$ or $\tup^\prime \neq \tup$ and $\repspnrw{A^\prime}(\doc)(\tup) =
  \repspnrw{A}(\doc)(\tup)$, otherwise. Furthermore, all runs $\rn \in
  \Rn{A^\prime}{\doc}$ have length $|\doc_\tup| + 1$. Therefore, the weight
  $\repspnrw{A^\prime}(\doc)(\tup)$ can be obtained by taking the sum of the
  weights of all runs of length $|\doc_\tup| + 1$ of $A^\prime$. If we assume,
  \mbox{w.l.o.g.}, that the states of $A^\prime$ are $\{1,\ldots,n\}$ for some
  $n \in \N$, then, due to distributivity of $\srplus$ over $\srtimes$, this sum
  can be computed as
  \[
    \repspnrw{A^\prime}(\doc)(\tup) = v_I \times (M_\delta)^{|\doc_\tup| + 1} \times
    (v_F)^T ,
  \]
  where
  \begin{itemize}
  \item $v_I$ is the vector $(I(1),\ldots,I(n))$,
  \item $M_\delta$ is the $n\times n$ matrix with $M_\delta(i,j) = \boplus_{a \in
      \Sigma \cup \varop{V}}\delta(i,a,j)$, and
  \item $(v_F)^T$ is the transpose of vector $v_F = (F(1),\ldots,F(n))$.
  \end{itemize}
  Therefore, by the assumption that $\srd$ has an efficient encoding
  (Definition~\ref{def:efficientSemiring}), the weight can be computed in
  polynomial time.
\end{proof}

\subsection{Best Weight Evaluation}
In many semirings, the domain is naturally ordered by some relation. For
instance, the domain of the probability semiring is $\Q^+$, which is ordered by
the $\leq$-relation. This motivates evaluation problems, where one is interested
in some kind of optimization of the weight. We start by giving the definition of
an ordered semiring.\footnote{Note that the following definition slightly
  deviates from the definition by Droste and Kuich~\cite{Droste09}. We require
  the order to be linear, as the maximal weight would otherwise not be well
  defined.}

\begin{defi}[{\normalfont similar to Droste and Kuich~\cite{Droste09}}]
  A commutative monoid $(\srd, \srplus, \srzero)$ is \emph{ordered} if it is
  equipped with a linear order $\linorder$ preserved by the $\srplus$
  operation. An ordered monoid is \emph{positively ordered} if $\srzero
  \linorder a$ for all $a \in \srd$. A semiring $\sr$ is \emph{(positively)
    ordered} if the additive monoid is (positively) ordered and multiplication
  with elements $\srzero \linorder a$ preserves the order.
\end{defi}
\noindent We consider the following two problems.

\smallskip
\algproblem{1.5cm}{\textsc{Threshold}}{Regular annotator $A$ over an ordered
  semiring, document $\doc \in \docs$, and a weight $w \in \K$.}{Is there a
  tuple $\tup$ with $w \linorder \repspnrk{A}{\K}(\doc)(\tup)$?}
\computeproblem{1.5cm}{\textsc{MaxTuple}}{Regular annotator $A$ over an
  ordered semiring and a document $\doc \in \docs$.}{Compute a tuple with
  maximal weight, if it exists.}
\smallskip

Notice that, if \textsc{MaxTuple} is efficiently solvable, then so is
\textsc{Threshold}. We therefore prove upper bounds for \textsc{MaxTuple} and
lower bounds for \textsc{Threshold}. The \textsc{Threshold} problem is sometimes
also called the \emph{emptiness problem} in the weighted automata literature. It
turns out that both problems are tractable for positively ordered semirings
that are bipotent (that is, for every $a,b \in \srd$ it holds that $a \srplus b
\in \{a,b\}$).

We first make an observation about positively ordered, bipotent semirings.
\begin{obs}\label{obs:linorderBipotenz}
  Let $\sr$ be a positively ordered, bipotent semiring. Then, for every $a,b \in
  \srd$ with $a\linorder b$ it holds that $a \srplus b = b$. 
\end{obs}

\begin{proof}
  As $\srd$ is positively ordered it holds that $\srzero \linorder a$. Thus,
  \[
    \srzero \srplus b \linorder a \srplus b\;,
  \]
  and therefore $b \linorder a \srplus b$. Due to $\srd$ being bipotent, $a
  \srplus b \in \{a,b\}$. Assume that $a \srplus b = a$. Thus, $b \linorder a$
  and it follows with $a \linorder b$ and antisymmetry of $\linorder$ that $a =
  b$. Therefore, $a \srplus b = b$ as claimed.
\end{proof}

Notice that Observation~\ref{obs:linorderBipotenz} shows that, for positively ordered, bipotent semirings, the $\srplus$-operator is the same as the $\max$ operator, where the maximum is taken over the linear order $\linorder$.
\begin{cor}\label{cor:linorderBipotenz}
  Let $\sr$ be a positively ordered, bipotent semiring. Then $\srplus = \max$.
\end{cor}

We prove our tractability result for positively ordered, bipotent semirings. Due to Corollary~\ref{cor:linorderBipotenz}, we can assume that $\srplus$ is the maximum operator over $\linorder$.
\begin{thm}\label{thm:maxWeightP}
  Let $\sr$ be a positively ordered semiring, where $\srplus = \max$ is the maximum operator over the linear order $\linorder$. Furthermore, let $A$ be
  a functional $\K$-weighted \vset-automaton, and let $\doc \in \docs$ be a
  document. Then \textsc{MaxTuple} for $A$ and $\doc$ can be solved in
  polynomial time.
\end{thm}
\begin{proof}
  By Proposition~\ref{prop:toExtendedConversion}, we can assume,
  \mbox{w.l.o.g.}, that $A$ is given as an extended functional $\K$-weighted
  \vset-automaton. Furthermore, as $\K$ is bipotent, it must hold that $a
  \srplus b \in \{a,b\}$ for every $a,b \in \srd$. Therefore, the weight of a
  tuple $t \in \repspnrk{A}{\K}(\doc)$ is always equal to the weight of one of
  the runs $\rn$ with $\tup = \tup_\rn$.

  Let $\rn \in \Rn{A}{\doc}$ be the run of $A$ on $\doc$ with
  maximal weight. Due to $\srplus = \max$, it must hold that
  $\weight{\tup_\rn} = \weight{\rn}$ and $\weight{\tup} \linorder
  \weight{\tup_\rn}$, as otherwise, $\rn$ would not be the run of $A$ on $\doc$
  with maximal weight. Therefore, in order to find the tuple with maximal weight, we
  need to find the run of $A$ on $\doc$ with maximal weight.

  To this end, we define a directed acyclic graph (DAG) which is obtained by
  taking a ``product'' between $A$ and $\doc$. Finding the run with the maximal
  weight then boils down to finding the path with maximal weight in this DAG\@.

  Assume that $A = (V,Q,I,F,\delta)$. Recall that $2^{\varop{V}}$ denotes the
  power set of $\varop{V}$. We define a weighted, edge-labeled DAG $G =
  (N,E,w)$, where each edge $e$ is in $N\times (\{\varepsilon\} \uplus
  (2^{\varop{V}} \times \{1,\ldots,|\doc|+1\})) \times N$ and $w$ assigns a
  weight $w(e) \in \K$ to every edge $e$. We note that an edge $(p,(T,i),q) \in
  E$ will encodes that a transition, labeled $T$, is reached after reading
  $d_{\mspan{1}{i}}$.

  More formally, let $N \df \{s,t\} \uplus \{(q,i) \mid q \in Q$ and $1 \leq i
  \leq |\doc|+1 \}$. We say that a node $n = (p,i)$ is in layer $i$ of $G$,
  where $s$ is in layer $0$ and $t$ in layer $|\doc|+2$. Furthermore, let $E$ be
  defined as follows:
  \begin{align*}
    E \df& \{(s,\varepsilon,(q,1)) \mid I(q) \neq \srzero\}\\
    \cup\ & \{((q,|\doc|+1),\varepsilon,t) \mid F(q) \neq \srzero\}\\
    \cup\ & \{((p,i),(T,i),(q,i)) \mid T \subseteq \varop{V} \text{ and } \delta(p,T,q) \neq
            \srzero\} \\
    \cup\ & \{((p,i),\varepsilon,(q,i+1)) \mid \doc_{\mspan{i}{i+1}} = a \text{ and } \delta(p,a,q) \neq
            \srzero\}\;.
  \end{align*}
  Furthermore, for $T \subseteq \varop{V}$ and $a \in \alphabet$, we define the
  weight $w(e)$ for all $e \in E$ as follows:
  \begin{align*}
    w((s,\varepsilon,(q,1))) &\df I(q)\\
    w(((q,|\doc|+1),\varepsilon,t)) &\df F(q) \\
    w(((p,i),(T,i),(q,i))) &\df \delta(p,T,q) \\
    w(((p,i),\varepsilon,(q,i+1))) &\df \delta(p,a,q)\;.
  \end{align*}

  Recall that, in extended weighted \vset-automata, the set of states $Q$ is a
  disjoint union of letter- and variable states, such that all transitions
  labeled by $\sigma \in \alphabet$ originate in letter states and all transitions
  labeled by $T \subseteq \varop{V}$ originate in variable states. Therefore,
  $G$ must be acyclic, as all edges are either from a node in layer $i$ to a
  node in layer $i+1$ or from a variable state to a letter state within the same
  layer. Furthermore, there is a path from $s$ to $t$ in $G$ with weight $w$ if
  and only if there is a tuple $\tup \in \repspnrk{A}{\K}(\doc)$ with the same
  weight.

  As shown in Mohri~\cite{Mohri2009}, a path of maximal weight can be computed
  in polynomial time. We also give procedure,
  Procedure~\ref{alg:best-weight-eval}, for the sake of
  succinctness.\footnote{We note that all semiring operations must be computable
    efficiently as we assume that only efficient encodings are used.} The
  correctness follows directly from $\srd$ being positively ordered, thus order
  being preserved by addition and multiplication with an element $\ell \in \K$.
\end{proof}

\begin{procedure}
  \DontPrintSemicolon
	\KwIn{A weighted, edge-labeled DAG $G=(N,E,w)$, nodes $s,t$}
  \KwOut{A path from $s$ to $t$ in $G$ with maximal weight or Null, if no such
    path exists.}
  $p(s) \gets \varepsilon$ \Comment*[r]{\textrm{$p(n)$ will store the
    best path from $s$ to $n$.}}
  $w(s) \gets \srone$ \Comment*[r]{\textrm{$w(n)$ will be the weight of the path
      $p(n)$.}}
  \For{$s \neq n \in N$ {\upshape in topological order}}{
    \If{{\upshape $n$ does not have an incoming edge in $E$, i.e., there is no
        $(n',\ell,n) \in E$}}
    {
      $p(n) = \varepsilon$\;
      $w(n) = \srzero$\;
    }\Else{
      $e = \argmax_{(n',\ell,n)\in E} w(n')\srtimes w((n',\ell,n))$\;
      $p(n) = p(n') \cdot e$\;
      $w(n) = w(n')\srtimes w(e)$\;
    }
  }
  \If{$w(t) \neq \srzero$}{
    \textbf{output $p(t)$}\;
  }\Else{
    \textbf{output Null}\;
  }
  \caption{\bestWeightEval(G,s,t)}\label{alg:best-weight-eval} 
\end{procedure}

If the semiring is not bipotent, however, the \textsc{Threshold} and
\textsc{MaxTuple} problems quickly become intractable. 
\begin{thm}\label{thm:maxWeightNp}
  Let $\sr$ be a semiring such that $\boplus_{i=1}^m \srone$ is
  strictly monotonously increasing for increasing values of
  $m$. Furthermore, let $A$ be a functional $\K$-weighted
  \vset-automaton, let $\doc \in \docs$ be a document, and $k \in \K$
  be a weight threshold. Then \textsc{Threshold} for such inputs is
  \NP-complete.
\end{thm}
\begin{proof}
  It is obvious that \textsc{Threshold} is in \NP, as one can guess a
  tuple $\tup$ and test in \ptime whether
  $w \linorder \repspnrk{A}{\K}(\doc)(\tup)$ using
  Theorem~\ref{thm:weightedTupleEvalPtime}.

  For the \NP-hardness, we will reduce from the MAX-3SAT problem. Given a 3CNF
  formula and a natural number $k$, the decision version of MAX-3SAT asks
  whether there is a valuation satisfying at least $k$ clauses. Let ${\psi = C_1
    \land \cdots \land C_m}$ be a Boolean formula in 3CNF over variables
  $x_1,\ldots,x_n$ such that each clause $C_i = (\ell_{i,1} \lor \ell_{i,2} \lor
  \ell_{i,3})$ is a disjunction of exactly three literals $\ell_{i,j} \in \{x_c,
  \lnot x_c \mid 1 \leq c \leq n\}$, $1 \leq i \leq k, 1 \leq j \leq 3$. We can
  assume, \mbox{w.l.o.g.}, that no clause has two literals corresponding to the
  same variable. Observe that, for each clause $C_i$, there are $2^3 = 8$
  assignments of the variables corresponding to the literals of $C_i$ of which
  exactly $7$ satisfy the clause $C_i$. Formally, let $f_{C_i}$ be the function
  that maps a variable assignment $\tau$ to a number between $1$ and $8$,
  depending on the assignments of the literals of the clause $C_i$. We assume,
  \mbox{w.l.o.g.}, that $f_{C_i}(\tau) = 8$ if and only if $C_i$ is not
  satisfied by $\tau$.

  We define a functional weighted automaton automaton $A_\psi$ over the unary
  alphabet $\Sigma = \{\sigma\}$ such that
  $\repspnrk{A_\psi}{\K}(\sigma^n)(\tup) = \boplus_{i=1}^m \srone$ if and only
  if the assignment corresponding to $\tup$ satisfies exactly $m$ clauses in
  $\psi$ and $\repspnrk{A_\psi}{\K}(\doc)(\tup) = \srzero$ if $\doc \neq
  \sigma^n$ or $\tup$ does not encode a variable assignment. To this end, each
  variable $x_i$ of $\psi$ is associated with a corresponding capture variable
  $x_i$ of $A_\psi$. With each assignment $\tau$ we associate a tuple
  $\tup_\tau$, such that
  \[
    \tup_\tau(x_i) =
    \begin{cases}
      \mspan{i}{i} & \text{ if } \tau(x_i) = 0, \text{ and}\\
      \mspan{i}{i+1} &\text{ if } \tau(x_i) = 1.\\
    \end{cases}
  \]
  The automaton $A_\psi \df (\Sigma, V,Q, I, F, \delta)$ consists of $m$ disjoint
  branches, where each branch corresponds to a clause of $\psi$; we call these
  \emph{clause branches}. Each clause branch is divided into $7$
  sub-branches, such that a path in the sub-branch $j$ corresponds to a variable
  assignment $\tau$ if $f_{C_i}(\tau) = j$. Thus, each clause branch has exactly one
  run $\rn$ with weight $\srone$ for each tuple $\tup_\tau$ associated to a
  satisfying assignment $\tau$ of $C_i$.

  More formally, the set of states $Q = \{ q_{i,j}^{a,b} \mid 1 \leq i \leq m, 1
  \leq j \leq n, 1 \leq a \leq 7, 1 \leq b \leq 5\}$ contains $5n$ states for
  every of the $7$ sub-branches of each clause branch. Intuitively, $A_\psi$ has
  a gadget, consisting of $5$ states, for each variable and each of the $7$
  satisfying assignments of each clause. Figure~\ref{fig:exGadget} depicts the
  three types of gadgets we use here. Note that the weights of the drawn edges
  are all $\srone$. We use the left gadget if $x$ does not occur in the relevant
  clause and the middle (resp., right) gadget if the literal $\lnot x$ (resp.,
  $x$) occurs. Furthermore, within the same sub-branch of $A_\psi$, the last
  state of each gadget is the same state as the start state of the next
  variable, i.e., $q_{i,j}^{a,5} = q_{i,j+1}^{a,1}$ for all $1 \leq i \leq k, 1
  \leq j < n, 1 \leq a \leq 7$.

  We illustrate the crucial part of the construction on an example. Let $\psi =
  (x_1 \lor \lnot x_2 \lor x_4) \land (x_2 \lor x_3 \lor x_4)$. The
  corresponding weighted \vset-automaton $A_\psi$ therefore has $14 = 2\times7$
  disjoint branches. Figure~\ref{fig:exBranch} depicts the sub-branch for clause
  $C_1$ that corresponds to all assignments with $x_1 = x_2 = 1$ and $x_4 = 0$.
  \begin{figure}
    \resizebox{\linewidth}{!}{
    \begin{tikzpicture}[>=latex]
      \def\w{2}
      \def\h{.75}
      \def\htwo{1.5}
      \node[state] (q1111) at (0,0) {$q_{1,1}^{1,1}$};
      \node[state] (q1112) at ($(q1111) + (\w,0)$) {$q_{1,1}^{1,2}$};
      \node[state] (q1113) at ($(q1112) + (\w,\h)$) {$q_{1,1}^{1,3}$};
      \node[state] (q1114) at ($(q1112) + (\w,-\h)$) {$q_{1,1}^{1,4}$};
      \node[state] (q1115) at ($(q1114) + (\w,\h)$) {$q_{1,1}^{1,5}$};
      \draw (q1111) edge[->] node[above,sloped] {$\vop{x_1}$}(q1112);
      \draw (q1112) edge[->] node[above,sloped] {$\vcl{x_1}$}(q1113);
      \draw (q1112) edge[->] node[below,sloped] {$\sigma$}(q1114);
      \draw (q1114) edge[->] node[below,sloped] {$\vcl{x_1}$}(q1115);
      \draw [decorate,decoration={brace,amplitude=10pt}]
        ($(q1111) + (0,\htwo)$) -- node[above=10pt] {$x_1$} ($(q1115) + (0,\htwo)$);
      
      \node[state] (q1212) at ($(q1115) + (\w,0)$) {$q_{1,2}^{1,2}$};
      \node[state] (q1213) at ($(q1212) + (\w,\h)$) {$q_{1,2}^{1,3}$};
      \node[state] (q1214) at ($(q1212) + (\w,-\h)$) {$q_{1,2}^{1,4}$};
      \node[state] (q1215) at ($(q1214) + (\w,\h)$) {$q_{1,2}^{1,5}$};
      \draw (q1115) edge[->] node[above,sloped] {$\vop{x_2}$}(q1212);
      \draw (q1212) edge[->] node[above,sloped] {$\vcl{x_2}$}(q1213);
      \draw (q1212) edge[->] node[below,sloped] {$\sigma$}(q1214);
      \draw (q1214) edge[->] node[below,sloped] {$\vcl{x_2}$}(q1215);
      \draw [decorate,decoration={brace,amplitude=10pt}]
        ($(q1215) + (0,-\htwo)$) -- node[below=10pt] {$x_2$} ($(q1115) + (0,-\htwo)$);
      
      \node[state] (q1312) at ($(q1215) + (\w,0)$) {$q_{1,3}^{1,2}$};
      \node[state] (q1313) at ($(q1312) + (\w,\h)$) {$q_{1,3}^{1,3}$};
      \node[state] (q1314) at ($(q1312) + (\w,-\h)$) {$q_{1,3}^{1,4}$};
      \node[state] (q1315) at ($(q1314) + (\w,\h)$) {$q_{1,3}^{1,5}$};
      \draw (q1215) edge[->] node[above,sloped] {$\vop{x_3}$}(q1312);
      \draw (q1312) edge[->] node[above,sloped] {$\vcl{x_3}$}(q1313);
      \draw (q1312) edge[->] node[below,sloped] {$\sigma$}(q1314);
      \draw (q1313) edge[->] node[above,sloped] {$\sigma$}(q1315);
      \draw (q1314) edge[->] node[below,sloped] {$\vcl{x_3}$}(q1315);
      \draw [decorate,decoration={brace,amplitude=10pt}]
        ($(q1215) + (0,+\htwo)$) -- node[above=10pt] {$x_3 \lor \lnot x_3$} ($(q1315) + (0,+\htwo)$);
      
      \node[state] (q1412) at ($(q1315) + (\w,0)$) {$q_{1,4}^{1,2}$};
      \node[state] (q1413) at ($(q1412) + (\w,\h)$) {$q_{1,4}^{1,3}$};
      \node[state] (q1414) at ($(q1412) + (\w,-\h)$){$q_{1,4}^{1,4}$};
      \node[state] (q1415) at ($(q1414) + (\w,\h)$){$q_{1,4}^{1,5}$};
      \draw (q1315) edge[->] node[above,sloped] {$\vop{x_4}$}(q1412);
      \draw (q1412) edge[->] node[above,sloped] {$\vcl{x_4}$}(q1413);
      \draw (q1413) edge[->] node[above,sloped] {$\sigma$}(q1415);
      \draw (q1414) edge[->] node[below,sloped] {$\vcl{x_4}$}(q1415);
      \draw [decorate,decoration={brace,amplitude=10pt}]
        ($(q1415) + (0,-\htwo)$) -- node[below=10pt] {$\lnot x_4$} ($(q1315) + (0,-\htwo)$);
      \end{tikzpicture}
    }
    \caption{The sub-branch of $A_\psi$ corresponding to $C_1$ and $x_1 = x_2 =
      1, x_4 = 0$.}\label{fig:exBranch}
  \end{figure}
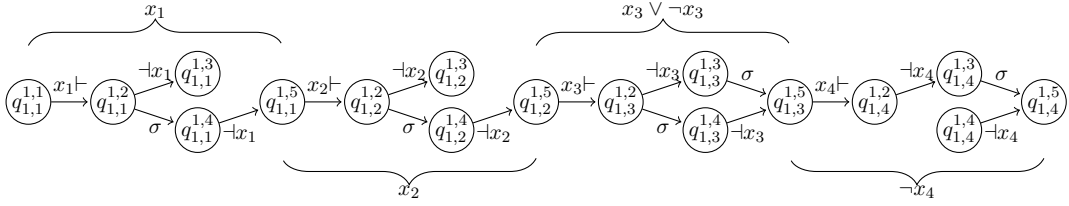
\begin{figure}
    \resizebox{\linewidth}{!}{
    \begin{tikzpicture}[>=latex, ->]
      \def\w{2}
      \def\h{.75}
      \def\wtwo{2}
      
      \node[state] (q1) at (0,0){};
      \node[state] (q2) at ($(q1) + (\w,0)$){};
      \node[state] (q3) at ($(q2) + (\w,\h)$){};
      \node[state] (q4) at ($(q2) + (\w,-\h)$){};
      \node[state] (q5) at ($(q4) + (\w,\h)$){};
      
      \node[state] (q11) at ($(q5) + (\wtwo,0)$){};
      \node[state] (q12) at ($(q11) + (\w,0)$){};
      \node[state] (q13) at ($(q12) + (\w,\h)$){};
      \node[state] (q14) at ($(q12) + (\w,-\h)$){};
      \node[state] (q15) at ($(q14) + (\w,\h)$){};
      
      \node[state] (q21) at ($(q15) + (\wtwo,0)$){};
      \node[state] (q22) at ($(q21) + (\w,0)$){};
      \node[state] (q23) at ($(q22) + (\w,\h)$){};
      \node[state] (q24) at ($(q22) + (\w,-\h)$){};
      \node[state] (q25) at ($(q24) + (\w,\h)$){};
      
      \draw (q1) edge node[above,sloped] {$\vop{x}$}(q2);
      \draw (q2) edge node[above,sloped] {$\vcl{x}$}(q3);
      \draw (q2) edge node[below,sloped] {$\sigma$}(q4);
      \draw (q3) edge node[above,sloped] {$\sigma$}(q5);
      \draw (q4) edge node[below,sloped] {$\vcl{x}$}(q5);
      
      \draw (q11) edge node[above,sloped] {$\vop{x}$}(q12);
      \draw (q12) edge node[above,sloped] {$\vcl{x}$}(q13);
      \draw (q13) edge node[above,sloped] {$\sigma$}(q15);
      \draw (q14) edge node[below,sloped] {$\vcl{x}$}(q15);
      
      \draw (q21) edge node[above,sloped] {$\vop{x}$}(q22);
      \draw (q22) edge node[above,sloped] {$\vcl{x}$}(q23);
      \draw (q22) edge node[below,sloped] {$\sigma$}(q24);
      \draw (q24) edge node[below,sloped] {$\vcl{x}$}(q25);
    \end{tikzpicture}
    }
    \caption{Example gadgets for variable $x$.}\label{fig:exGadget}
  \end{figure}
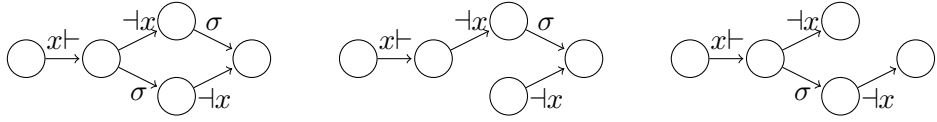
  
  Formally, the initial weight function is $I(q_{i,j}^{a,b}) = \srone$ if $j = 1
  = b$ and $I(q_{i,j}^{a,b}) = \srzero$ otherwise. The final weight function
  $F(q_{i,j}^{a,b}) = \srone$ if $j = n$ and $b = 5$ and $F(q_{i,j}^{a,b}) =
  \srzero$, otherwise. The transition function $\delta$ is defined as follows:
  \[
    \delta(q_{i,j}^{a,b}, o, q_{i,j}^{a,b'}) =
    \begin{cases}
      \srone  & b = 1, b' = 2, o = \vop{x_j} \\
      \srone  & b = 2, b' = 3, o = \vcl{x_j} \\
      \srone  & b = 2, b' = 4, o = \sigma, \text{ and there is a variable assignment
        $\tau$ such that} \\
        &\text{ $\tau(x_j) = 1$ and $f_{C_i}(\tau) = a$} \\ 
      \srone  & b = 3, b' = 5, o = \sigma, \text{ and there is a variable assignment
        $\tau$ such that} \\
        &\text{ $\tau(x_j) = 0$ and $f_{C_i}(\tau) = a$} \\
      \srone  & b = 4, b' = 5, o = \vcl{x_j} \\
    \end{cases}
  \]
  All other transitions have weight $\srzero$.

  We show that there is a tuple $\tup \in \repspnrk{A_\psi}{\K}(\sigma^n)$ with
  weight $\weight{\tup} = \boplus_{i=1}^k \srone$ if and only if the
  corresponding assignment $\tau$ satisfies exactly $k$ clauses of $\psi$. To
  this end, let $\tau$ be an assignment of the variables $x_1,\ldots,x_n$. Thus,
  there is a run $\rn \in \Rn{A_\psi}{\sigma^n}$ with weight $\weight{\rn} =
  \srone$ starting in $q_{i,0}^{a,0}$, such that $a = f_{C_i}(\tau)$ if and only
  if $\tau$ satisfies clause $C_i$. Due to $\boplus_{i=1}^k \srone$ being
  strictly monotonously increasing it follows that $\boplus_{i=1}^k \srone
  \linorder \weight{\tup_\tau}$ if and only if the corresponding assignment to
  $\tau$ satisfies at least $k$ clauses. Let $w = \boplus_{i=1}^k \srone$. It
  follows directly that there is an assignment $\tau$ of $\psi$ satisfying $k$
  clauses if and only if there is a tuple $\tup$ with $w\linorder
  \repspnrk{A_\psi}{\K}(\sigma^n)(\tup)$.
\end{proof}

We note that Theorem~\ref{thm:maxWeightP} and Theorem~\ref{thm:maxWeightNp} give
us tight bounds for all semirings we defined in Example~\ref{ex:semirings}.

Since MAX-3SAT is hard to approximate, we can turn Theorem~\ref{thm:maxWeightNp}
into an even stronger inapproximability result for semirings where approximation
makes sense. To this end, we focus on semirings that contain $(\N,+,\cdot,0,1)$
as a subsemiring in the following result. Note that this already implies that
$\boplus_{i=1}^m \srone$ is strictly monotonously increasing for increasing
values of $m$.

\begin{thm}\label{thm:MaxTupleNoApprox}
  Let $\K$ be a semiring that contains $(\N,+,\cdot,0,1)$ as a subsemiring and
  let $A$ be a weighted \vset-automaton over $\K$. Unless \ptime = \NP, there is
  no algorithm that approximates the tuple with the best weight within a
  sub-exponential factor in polynomial time.
\end{thm}
\begin{proof} 
  Given a Boolean formula $\psi$ in 3CNF, MAX-3SAT asks for the maximal number
  of clauses satisfied by a variable valuation. Given a 3CNF formula, a
  polynomial time approximation algorithm with approximation factor $x$ returns
  a variable assignment satisfying at least $\frac{\text{opt}}{x}$ clauses,
  where $\text{opt}$ is the maximal number of clauses which are satisfiable by a
  single variable assignment. H{\aa}stad~\cite{Hastad01} showed that, for every
  $\varepsilon > 0$, it is \NP-hard to approximate MAX-3SAT with an
  approximation factor $x < 8/7 - \varepsilon$, even if the input is restricted
  to satisfiable 3CNF instances. In other words, unless \ptime = \NP, for every polynomial time approximation algorithm, there is a
  satisfiable 3CNF instance such that, for this instance, the algorithm does not return an approximation
  with approximation factor $x < 8/7 - \varepsilon$. We can leverage this, using
  the reduction from Theorem~\ref{thm:maxWeightNp}, to show that there is no
  polynomial time algorithm that approximates the tuple with the best weight
  with a sub-exponential approximation factor.
  
  Let $\psi$ be a satisfiable 3CNF formula with $m$ clauses and let $A_\psi$ be
  the weighted \vset-automaton and $\doc \in \docs$ be as constructed from
  $\psi$ in the proof of Theorem~\ref{thm:maxWeightNp}. Let $c$ be the size of
  $A_\psi$, which is linear in $n$. As shown in Theorem~\ref{thm:maxWeightNp},
  there is a tuple $\tup$ in $A_\psi$ with weight $j$ if and only if the
  variable assignment corresponding to $\tup$ satisfies exactly $j$ clauses. For
  a $k \in \N$ let $A_\psi^k$ be the weighted \vset-automaton, constructed by
  concatenating $k$ fresh copies of $A_\psi$, each of which operates on a set of
  $n$ fresh variables, by inserting $\varepsilon$-edges with
  weight $\srone$ from $q_i$ to $q_{i+1}$ where $q_i$ is a final state of the
  $i$-th copy and $q_{i+1}$ an initial state of the $i+1$-th copy. Observe that
  $A_\psi^k$ has size $c\cdot k$, has $nk$ variables, and each tuple $\tup \in
  \repspnrk{A_\psi^k}{\K}(d^{k})$ encodes $k$, possibly different, variable
  assignments for $\psi$.

  For the sake of contradiction, assume there is a polynomial time algorithm
  approximating the tuple with the best weight within a sub-exponential factor
  $f(x) \in o(b^x)$ for every $b > 1$ (i.e., for every $b > 1$ and every $C > 0$
  there is a $x_0 > 0$, such that $f(x) < C \cdot b^x$, for every $x > x_0$).
  That is, given a spanner $A$ of size $c$ and a document $\doc$ of size $|\doc|
  \leq c$, the approximation algorithm returns a tuple $\tup$ with
  $\weight{\tup} \geq \frac{\text{opt}}{f(c)}$, where $\text{opt}$ is the
  maximal weight assigned to a tuple $\tup$ over $\doc$ by $A$.

  Due to H{\aa}stad~\cite[Theorem 6.5]{Hastad01}, there is a satisfiable 3CNF
  formula $\psi$, such that this procedure can at best lead to an
  $(8/7-\varepsilon)$ approximation of the maximal number of satisfiable
  clauses. Therefore, it follows that
  \[
    \weight{\tup} \leq \frac{\text{opt}}{(8/7 - \varepsilon)^k}\;.
  \]

  Let $\tup \in \repspnrk{A_\psi^k}{\K}(d^{k})$ be such an approximation and
  $\tau_1,\ldots,\tau_k$ be the corresponding variable assignments of $\psi$.
  Recall that $|A_\psi^k| = c \cdot k$ and $|d^{k}| = n \cdot k \leq c \cdot k$.
  Per assumption, there is an approximation algorithm, returning a tuple $\tup$
  with
  \[
    \weight{\tup} \geq \frac{\text{opt}}{f(c\cdot k)}\;.
  \]
  The tuple $\tup$ encodes $k$ variable assignments and the weight of the tuple
  is the product of the weights of the variable assignments. Let $\tau$ be one
  of the variable assignments, encoded by $\tup$, which satisfy the most
  clauses.\footnote{We note that there might be multiple assignments satisfying
    the same number of clauses.}

  Thus, combining both inequalities, it must hold that
  $\frac{\text{opt}}{f(c\cdot k)} \leq \weight{\tup} \leq \frac{\text{opt}}{(8/7
    - \varepsilon)^k}$. Thus, $(8/7-\varepsilon)^k \leq f(c\cdot k)$. Recall
  that the function $f$ is sub-exponential, that is $f(x) \in o(b^x)$ for every $b >
  1$. Therefore, $(8/7-\varepsilon)^k \in o(b^{(c\cdot k)})$, however, if $1 < b <
  8/7-\varepsilon$, this does not hold for arbitrarily large $k$, leading to the
  desired contradiction.
\end{proof}
 
\section{Enumeration Problems}\label{sec:enum}
In this section we consider computing the output of annotators from
the perspective of enumeration problems, where we try to enumerate all
tuples with nonzero weight, possibly from large to small. Such
problems are highly relevant for (variants of) \vset-automata, as
witnessed by the recent literature that explicitly focuses on such
automata~\cite{AmarilliBMN19,FlorenzanoRUVV18} or on alternative
formalisms~\cite{BourhisGJR21,DeepK21}.

An \emph{enumeration problem} $P$ is a (partial) function that maps each input
$i$ to a finite or countably infinite set of \emph{outputs for $i$}, denoted by
$P(i)$. Terminologically, we say that, given $i$, the task is to \emph{enumerate
  $P(i)$}.

An \emph{enumeration algorithm} for $P$ is an algorithm that, given
input $i$, writes a sequence of answers to the output such that every
answer in $P(i)$ is written precisely once. If $A$ is an enumeration
algorithm for an enumeration problem $P$, we say that $A$ runs in
preprocessing $p$ and delay $d$ if the time before writing the first
answer is $p(|i|)$, where $|i|$ is the size of the input $i$, and the
time between writing every two consecutive answers is $d(|i|)$. By
\emph{between answers}, we mean the number of steps between writing
the first symbol from an answer until writing the first symbol of the
next answer.  We generalize this terminology in the usual way to
classes of functions.  E.g., an algorithm with linear preprocessing
and constant delay has a linear function for $p$ and a constant
function for $d$. Notice that we measure with a RAM model with
logarithmic word size. Therefore, for a document $d$ and
\vset-automaton $A$, an algorithm with a delay that is constant in
$|d|$ and polynomial in $|A|$ is able to produce answers that consist
of $|A|^c$ many words of logarithmic size, for some constant
$c$. Thus, such an enumeration algorithm can produce $\doc$-tuples
$\tup \in \repspnr{A}(\doc)$ as answers.

Given a $\K$-weighted \vset-automaton $A$ and a document $\doc$, let $f(A,\doc)$
be the maximal time required for a single addition or multiplication while
computing the weight $\repspnrw{A}(\doc)(\tup)$ for some tuple $\tup$. We note
that, due to the assumption that $\K$ has an efficient encoding, $f(A,\doc)$ is
at most polynomial in $|A|$ and $|\doc|$. Furthermore, for instance for finite
semirings (like the Boolean semiring or the access control semiring),
$f(A,\doc)$ is constant. If the order of the answers does not matter and the
semiring is positive, we can guarantee an enumeration algorithm which has linear
preprocessing time and constant delay in the size of the document and polynomial
time and delay in the size of $A$ and $f(A,\doc)$.\footnote{We note that
  $f(A,\doc)$ can be polynomial in the size of the document. Thus, strictly
  speaking, preprocessing (resp., delay) might not be linear (resp., constant)
  in the size of the document. However, stating the theorem like this enables us
  to give two direct corollaries (Corollaries~\ref{cor:fconstenum}
  and~\ref{cor:fpolyenum}) depending on whether or not $f(A,\doc)$ is constant.}
Note that the proof of the theorem essentially requires to go through the entire
proof of the main result of Amarilli et al.~\cite[Theorem
1.1]{AmarilliBMN19}.

\begin{thm}\label{thm:constantDelayEnum}
  Given a weighted functional \vset-automaton $A$ without $\varepsilon$-transitions over a positive semiring $\K$,
  and a document $\doc$, the $\K$-Relation $\repspnrk{A}{\K}(\doc)$ can be
  enumerated with preprocessing linear in $|\doc|$ and polynomial in $|A|$ and
  $f(A,\doc)$, and delay constant in $|\doc|$ and polynomial in $|A|$ and
  $f(A,\doc)$.
\end{thm}
\begin{proof}[Proof Sketch]
  Amarilli et al.~\cite[Theorem 1.1]{AmarilliBMN19} showed that, given a
  sequential \vset-automaton $A$ 
   without $\varepsilon$-transitions
  and a document $\doc$, one can enumerate
  $\repspnr{A}(\doc)$ with preprocessing time $O((|Q|^{\omega+1}+ |A|) \times
  |\doc|)$ and with delay $O(|V| \times (|Q|^2 + |A| \times |V|^2))$, where $2
  \leq \omega \leq 3$ is an exponent for matrix multiplication, $V$ is the set
  of variables, and $Q$ the set of states in $A$. In other words,
  $\repspnr{A}(\doc)$ can be enumerated with linear preprocessing and constant
  delay in $\doc$, and polynomial preprocessing and delay in $A$. To obtain this
  result, they view the transition function of $A$ as a (Boolean) transition
  matrix. Their methods easily extend from the Boolean case to transition
  matrices over positive semirings.\footnote{Note that positivity is required as
    otherwise weights might sum up or multiply to zero, which may violate the
    constant delay.} The claimed complexity for enumeration of the $\K$-Relation
  $\repspnrk{A}{\K}(\doc)$ can be achieved by computing all matrix
  multiplications over $\K$ instead of $\B$. Furthermore, instead of storing the
  set $\Lambda$ of current states, one has to store a set of
  (state,weight)-tuples in order to compute the correct weights of the returned
  tuples.
\end{proof}

Depending on whether or not $f(A,\doc)$ is constant, we have the following two
corollaries.

\begin{cor}\label{cor:fconstenum}
  Given a weighted functional \vset-automaton $A$ over a positive semiring $\K$,
  and a document $\doc$, such that $f(A,\doc)$ is constant. Then the
  $\K$-Relation $\repspnrk{A}{\K}(\doc)$ can be enumerated with preprocessing
  linear in $|\doc|$ and polynomial in $|A|$, and delay constant in $|\doc|$ and
  polynomial in $|A|$.
\end{cor}

\begin{cor}\label{cor:fpolyenum}
  Given a weighted functional \vset-automaton $A$ over a positive semiring $\K$,
  and a document $\doc$, such that $f(A,\doc)$ is polynomial in $|A|$ and
  $|\doc|$. Then the $\K$-Relation $\repspnrk{A}{\K}(\doc)$ can be enumerated
  with preprocessing linear in $|\doc|$ and polynomial in $f(A,\doc)$, and delay
  constant in $|\doc|$ and polynomial in $f(A,\doc)$.
\end{cor}

Thus, if the order of the answers does not matter and the semiring is positive,
we can guarantee an efficient enumeration algorithm.

We now consider cases in which answers are required to arrive in a certain
ordering.

\smallskip
\computeproblem{3cm}{Ranked Annotator Enumeration
    \textsc{(\rase)}}{Regular annotator $A$ over an ordered
  semiring $\sr$ and a document $\doc$.}{Enumerate all tuples $\tup \in
  \repspnrk{A}{\K}(\doc)$ in descending order on $\K$.}  
\smallskip

\begin{thm}\label{thm:bipotentRankedEnum}
  Let $\srd$ be an positively ordered, bipotent semiring, let $A$ be a
  $\srd$-weighted functional \vset-automaton, and let $\doc \in \docs$ be a
  document. Then \rase can be solved with polynomial delay and preprocessing.
\end{thm}
\begin{proof}
  By Proposition~\ref{prop:toExtendedConversion}, we can assume that $A$ is an
  extended functional $\K$-weighted \vset-automaton. Therefore, all runs of $A$
  which accept a tuple $\tup\in\repspnrk{A}{\K}(\doc)$ have the same label.
  We will use the DAG $G$ we defined in the proof of
  Theorem~\ref{thm:maxWeightP} and run a slight adaptation of Yen's
  algorithm~\cite{Yen71} on $G$. 
   
  From the proof of Theorem~\ref{thm:maxWeightP} it follows that, given $G$ we
  can find a path from $s$ to $t$ in $G$ with maximal weight in polynomial time.
  Let $p = n_0\cdot e_0 \cdot n_1 \cdots e_{k-1}\cdot n_k$ be a path in $G$,
  where $n_i \in N$ and $e_j \in E$ for $0 \leq i \leq k$ and $0 \leq j < k$. We
  denote by $p[i,i]$ the node $n_i$, by $p[i,j]$ the path $n_i\cdot e_i \cdots
  e_{j-1}\cdot n_j$ and by $N(p)$ the set $\{n_i \mid 0 \leq i \leq k\}$ of
  nodes used by $p$. The Procedure~\ref{alg:yen-label} shows how Yen's 
  algorithm can be adapted for the \rase problem. Recall that per construction
  of $D$, all edges which correspond to variable edges of $A$, are labeled by a
  tuple $(T,i)$, which encodes that the set $T$ is processed after reading
  $\doc_{\mspan{1}{i}}$. Thus, line~\ref{alg:yen:delell} ensures that,
  whenever the algorithm reaches line~\ref{alg:yen:gprime}, all paths $p[0,i]
  \cdot p'$ where $p'$ is a path from $p[i,i]$ to $t$ in $G'$ differ from the
  paths in the set $\text{Out}$ in at least one edge label and therefore, no
  tuple is enumerated multiple times. Observe that the first output of
  Procedure~\ref{alg:yen-label} is generated after polynomial time. Furthermore,
  every iteration of the while loop line~\ref{alg:yen:while}, takes polynomial
  time. Thus, the algorithm runs with polynomial preprocessing and delay.
\end{proof}

\begin{procedure}
\DontPrintSemicolon	
\KwIn{A weighted, edge-labeled DAG $G=(N,E,w)$, as constructed in
  Theorem~\ref{thm:bipotentRankedEnum}, nodes $s,t$}
\KwOut{All paths from $s$ to $t$ in $G$ in decreasing order without repetitions
  of the same path labels.}
$\text{Out}$ $\gets$ $\emptyset$ \Comment*[r]{\textrm{$\text{Out}$ is the set of paths already
    written to output.}}
$\text{Cand}$ $\gets$ $\emptyset$ \Comment*[r]{\textrm{$\text{Cand}$ is a set of candidate paths
    from $s$ to $t$.}} 
$p$ $\gets$ \ref{alg:best-weight-eval}$(G,s,t)$\; 
\While{$p \neq$ \normalfont Null}{\label{alg:yen:while}
\textbf{output} $p$\; 
Add $p$ to $\text{Out}$ \;
\For{$i = 0$ {\upshape to} $|p|-1$\label{alg:yen:7}} {  
  $G' \gets (N, E', w)$, {\upshape where $E'$ is a copy of $E$}\; 
  \For{{\upshape every path $p_1$ in $\text{Out}$ with $p_1[0,i] =
      p[0,i]$}}{
    $n_i \cdot (n_i,\ell_i,n_{i+1}) \cdot n_{i+1} \gets p_1[i,i+1]$\;
    \For{{\upshape every $p,q \in N$ with $(p,\ell_i,q) \in E'$}}{
      Remove the edge $\big(n_i,\ell_i,n)$ from $E'$\Comment*[r]{\textrm{Delete all $\ell_i$-labeled
          edges.}}\label{alg:yen:delell} 
    }
  }
  $p_2 \gets$ \ref{alg:best-weight-eval}$(G',p[i,i],t)$\;\label{alg:yen:gprime}
  \If{\upshape$p_2$ is not Null}{
    Add $p[0,i]\cdot p_2$ to $\text{Cand}$\;
  }
}
$p \gets$ a path in $\text{Cand}$ with maximal weight \Comment*[r]{\textrm{$p \gets$ Null
    if $\text{Cand} = \emptyset$.}}
Remove $p$ from $\text{Cand}$
}
\caption{\bestWeightEnum(G,s,t).}
\label{alg:yen-label} 
\end{procedure}  
 
\section{Concluding Remarks}\label{sec:conclusion}
We embarked on a study that incorporates annotation, or weights, in
information extraction.  We proposed $\K$-annotators as a candidate
formalism to study this problem. The $\K$-annotators can be
instantiated with weighted \vset-automata, thereby obtaining the
regular $\K$-annotators.  We showed tha the regular $\K$-annotators
have favorable closure properties, such as closure under union,
projection, natural join, and, depending on the semiring, also under
string selection using regular relations.  The first complexity
results on evaluation problems are encouraging: answer testing is
tractable and, depending on the semiring, problems such as the
threshold problem, the max tuple problem, and enumeration of answers
are tractable too.

We note that the addition of weights to \vset-automata also introduces
new challenges. For instance, some questions which are typically
studied in database theory are not yet fully understood for weighted
automata, which are the basis of weighted \vset-automata. Examples are
equivalence and emptiness. Concerning equivalence, it is known that
equivalence is undecidable for weighted \vset-automata over the
tropical semiring (c.f.  Proposition~\ref{prop:equivalenceUndec}). In
general, however, it is not completely clear for which semirings
equivalence is decidable or not. Concerning emptiness, the definition
in weighted automata literature is not as database theoreticians would
expect. That is, it does not ask if there exists a document $\doc$
such that the automaton returns at least one tuple with nonzero weight
on $\doc$, but is additionally given a threshold (as in our
\textsc{Threshold} problem) and asks if the automaton returns a tuple
with at least the threshold weight (which requires an order on the
semiring). It is not yet clear how much this threshold influences the
complexity of the problem.

An additional challenge is that \emph{determinization} of weighted
automata is a complex matter and not always possible. It is well-known
to be possible for the Boolean semiring but, for the tropical semiring
(defined as $(\Q \cup \{-\infty\}, \max, +, -\infty, 0)$)
deterministic weighted automata are strictly less expressive than
unambiguous weighted automata, which are strictly less expressive than
general weighted automata, cf.~Klimann et al.~\cite{KlimannLMP04}.

An interesting direction for future research is to determine the exact
connection between our work and past work on Datalog with weights or
provenance. As we explained in Section~\ref{sec:datalog}, one can
transform a weighted vset-automaton into a Datalog program over
annotated relations according to the semantics of Deutch et
al.~\cite{DBLP:conf/icdt/DeutchMRT14}.  There are other notions of
weighted Datalog where the connection to our work appears to be
weaker, such as the \e{Independent Choice
  Logic}~\cite{DBLP:conf/uai/Poole95},
\e{Prism}~\cite{DBLP:conf/ijcai/SatoK97} and
\e{ProbLog}~\cite{DBLP:conf/ijcai/RaedtKT07} where the existences of
ground rules are probabilistically independent events, and the
\e{generative
  Datalog}~\cite{DBLP:journals/tods/BaranyCKOV17,DBLP:conf/pods/GroheKKL20}
where rules are able to produce random numbers. Yet, as far as we
know, the publications on weighted versions of Datalog include neither
complexity nor closure results that seem to imply the ones of this
article.  In particular, the tractability of the ``provenance
circuits'' of Deutch et al.~\cite{DBLP:conf/icdt/DeutchMRT14} applies
to special cases of semirings (e.g., \e{absorptive}) that do not arise
in our work. Whether the circuit approach can lead to new algorithms
or simplification of our results remains to be studied.

\section*{Acknowledgment}
\noindent The authors are grateful to Matthias Niewerth for many useful
discussions and his help regarding Theorem~\ref{thm:constantDelayEnum} and
Shaull Almagor for many helpful comments regarding weighted automata.
Furthermore, we thank the anonymous reviewers of ICDT 2020 and LMCS for many helpful
remarks. Johannes Doleschal is supported by the German Israeli Foundation (GIF), Grant l-1502-407.6/2019. 
Benny Kimelfeld is supported by the
DIP program of the 
Deutsche
Forschungsgemeinschaft (DFG, German Research Foundation), Grant 412400621, and the Israel Science Foundation (ISF), Grant 768/19.
Wim Martens is supported by the Deutsche
Forschungsgemeinschaft, Grant 431183758. A part of
the work of Liat Peterfreund was done while she was affiliated with the
Technion, CNRS, IRIF --- Université Paris Diderot supported by Fondation Sciences Math\'{e}matiques de Paris (FSMP), and University of Edinburgh.

\bibliographystyle{alphaurl}
\bibliography{references}

\end{document}